\documentclass[12pt,a4paper]{article}

\usepackage{tikz}
\usetikzlibrary{arrows.meta}

\usepackage[
  margin=0.75in,
  headsep=10pt    % separation between header rule and text
]{geometry}

\usepackage{graphicx}% Include figure files
\usepackage{epstopdf}   % eps -> pdf
\usepackage{amsmath}
\usepackage{amssymb}
\usepackage{breqn}      % automatic line breaks in maths
\usepackage{makecell}
\usepackage{multirow}
\usepackage{booktabs}
\usepackage{siunitx}
\usepackage{dcolumn}
\usepackage{bm}
\usepackage{xcolor}
\usepackage{textcomp}
\usepackage{gensymb}
\usepackage{textcomp}
\usepackage{subcaption}
\usepackage{lineno}
\usepackage{etoolbox}
\usepackage{authblk}
\usepackage{setspace}
\usepackage{float}
\usepackage{times}      % Times text + math (with mathptmx)
\usepackage[utf8]{inputenc}
\usepackage[T1]{fontenc}
\usepackage{soul} % For highlighting
\graphicspath{ {./figures2D/} }

\usepackage[sort&compress]{natbib}
\usepackage[hidelinks]{hyperref}

\title{Comparative Investigations on Active and Passive Tails of Undulating Swimmers}

\author[1,2]{Dev Pradeepkumar Nayak$^{\dag}$}
\author[1]{Ali Tarokh}
\author[2,$^{\dag}$]{Muhammad Saif Ullah Khalid}

\affil[1]{Department of Mechanical and Mechatronics Engineering, Lakehead University, Thunder Bay, Canada}
\affil[2]{Nature-Inspired Engineering Research Lab, Department of Mechanical and Mechatronics Engineering, Lakehead University, Thunder Bay, Canada}
\affil[*]{\textit{Corresponding author:} mkhalid7@lakeheadu.ca}

\begin{document}
\maketitle

%%%%%%%%%%%%%%%%%%%%%%%%%%%%%%%%%%%%%%%%%%%%%%%%%%%%%%%%%%%%%%%%%%%%%%%%%%%%%%%
\begin{abstract}
Fish display remarkable swimming capabilities through the coordinated interaction of the body and caudal fin, yet the potential role of a passively pitching tail in enhancing hydrodynamic performance remains unresolved. In this work, we evaluate the performance of a carangiform swimmer equipped with either an actively pitching tail or a passively pitching tail. High-fidelity fluid–structure interaction simulations are employed to assess how variations in joint stiffness, damping, and inertia influence thrust generation, power demand, and overall stability at two representative Reynolds numbers, $500$ and $5000$. The results reveal that actively pitching tails tend to generate greater thrust, while passively pitching tails deliver improved outcomes in terms of power demand at the lower Reynolds number. Larger pitching amplitudes contribute positively only when associated with higher swimming frequency; when produced by reduced inertia or more flexible joints, they lead to unfavorable effects. At the higher Reynolds number, active tails consistently outperform passive ones, although a small subset of passive cases still achieve favorable performance. Across all cases, a recurring balance emerges, with thrust production and power expenditure varying inversely. These findings clarify the hydrodynamic consequences of passive versus active tail motion and establish design principles for bio-inspired underwater vehicles, where smaller swimmers may benefit from passive tail pitching, while larger swimmers are better served by active control.
\end{abstract}

\section{INTRODUCTION}

Since a long time, researchers look at nature for inspiration, leading to the exploration and unlocking of many previously uncharted areas in engineering. Drawing insights from the underwater world, aquatic creatures demonstrate a remarkable balance of maneuverability, control, efficiency, adaptability, and speed. This natural efficiency continues to inspire the development of robotic technologies for both aerial and aquatic environments. Through the evolutionary process of natural selection, fish adapt to their surroundings and emerge as exceptional swimmers. Their undulatory motion enables them to achieve impressive propulsion and agile maneuverability. The interaction between the fish body and the surrounding fluid was studied in the past through experimental approaches \cite{thandiackal2023line, esposito2012robotic, flammang2011volumetric, marras2012fish} as well as using computational fluid dynamics (CFD) simulations \cite{dong2021fish, ming20193d, fardi2025characterizing}.

\begin{figure}[ht!]
    \centering
    \includegraphics[width=1\linewidth]{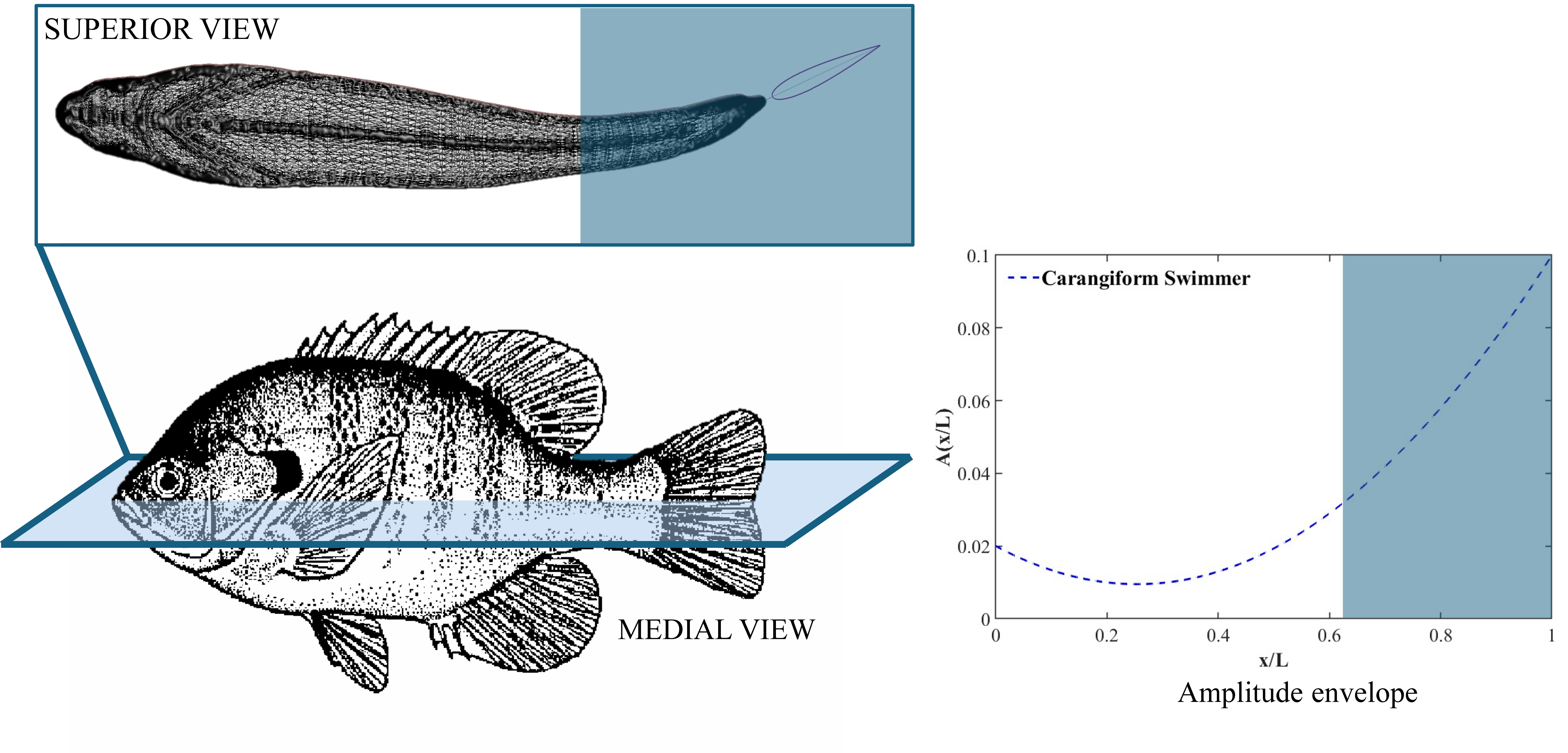}
    \caption{Medial view of a  Bluegill sunfish \cite{lauder2000function} and the superior view for its $2D$ Kinematics.}
    \label{fig:myMedialFish}
\end{figure}

Nature perfects the art of efficient swimming, with countless aquatic species exhibiting intricate body-fins coordination to navigate water with remarkable ease. This observation has inspired numerous studies in biomimetic propulsion in the recent past. Various factors, including climate change and pollution, continue to affect oceanic ecosystems \cite{mengerink2014call}, making the study of marine life and its adaptive locomotion increasingly important. Drawing inspiration from fish locomotion and their maneuvering techniques presents significant opportunities for the advancement of autonomous Underwater Vehicles (AUVs) \cite{amal2024bioinspiration}. Underwater exploration systems benefit from biologically inspired navigation strategies, as demonstrated in recent studies on odor dynamics that may guide robots in adapting to their environment \cite{kamran2024does}. Moreover, the integration of computational methods enables detailed exploration of fish behaviors using techniques, such as deep recurrent Q networks (DRQN) combined with the Immersed Boundary–Lattice Boltzmann Method (IB–LBM), optimizing tasks like Kármán gaiting and prey capture \cite{zhu2021numerical}. These advancements mark a critical step in marine robotics and underwater innovation. The undulatory kinematics of fish help them maneuver and swim efficiently underwater. Carangiform swimmers show relatively larger displacements in the posterior end of the body, with undulatory motion concentrated near the peduncle and the caudal fin. Numerous studies discuss the importance of the shape and size of the peduncle and the caudal fin on fish locomotion \cite{lauder2000function, hang2022active}. In Fig.~\ref{fig:myMedialFish}, the superior view of a Bluegill sunfish is shown to demonstrate the flexion of the body in its undulatory state. The amplitude envelope on the right shows that the maximum displacement occurs toward the posterior end of the fish exhibiting a typical carangiform swimming mode.

In recent years, multiple experimental and numerical investigations suggested that passive foils were a remarkable choice for mimicking freely swimming fish capable of producing thrust and propelling themselves \cite{lauder2012passive, feilich2015passive}. As reported by Lauder et al.~\cite{lauder2012passive}, flexible foils emulating the caudal fin and compared at constant stiffness exhibited large deviations in self-propelled speeds depending on the shape of the trailing edge. This finding highlights the importance of morphological variation in caudal fins among different fish species, tailored to their specific locomotion need. The study also suggested that improved thrust and swimming speeds could be predicted using further optimizing the passive foil models with active stiffness control. In contrast. Liu et al \cite{liu2019image} performed a coupled finite-element-method (FEM)-based computational structural dynamics model and an immersed-boundary-method (IBM)-based computational fluid dynamics (CFD) solver to inversely determine in vivo material properties and model flow-structure interactions between fins of fish. The study could predict how fin moves, and contribute to producing thrust. In their recent works, Hang et al \cite{hang2022active} reflected on the joint characteristic of the caudal fin with the fish's body exhibiting prominent flexion closer to the posterior end. Their work discussed the effects of both the active and passive bending of the body on the swimming performance. The conclusion was drawn in favor of passive flexion in terms of swimming efficiency but against it in terms of the swimming speed. Fish and Lauder \cite{fish2017control} discussed the extended operations of fins to control surfaces and not only propulsive structures, their investigations sets the stage to show how the active control and the passive control plays a role in both maneuverability and swimming performance. Behbahani et al \cite{behbahani2016design} introduced a flexible passive (feathering) joint for pectoral fins of a robotic fish that let the fin sweep back during the recovery stroke to cut drag while preserving prescribed rowing motion in the power stroke. Using this strategy, complexity of the robotic framework was minimized, and drag was reduced from the passive control. 

A Novel approach introduced by Qiu et al \cite{qiu2022locomotion} demonstrated a novel tendon-driven robotic fish with an active tail and a passively moving caudal fin, having a variable stiffness. This technique improved the steering radius and increased the swimming velocity. Similar observation were reported by Chen et al \cite{chen2020exploration} where the compliment joint with two identical torsional spring was proposed for a multi-joint robotic fish. The concept of a passive tail was further extended to a modular adaptive variable-stiffness passive joint (MAVSPJ) for a robotic fish recently proposed by Wang et al \cite {wang2025design}, where the swimming performance was improved with adaptively changing the stiffness of a torsional spring. Lu et al \cite{lu2024effect} designed and fabricated the pectoral fin based on a Fin Ray with a capability of hybrid active and passive deformations (APD). They reported an increase of $61.5\%$ thrust under the optimal condition compared to a traditional passively deformed pectoral fin. Some previous studies also highlighted the benefits of using a passive joint between the body and fins, whether pectoral or caudal. While they observed its positive effects, the underlying reason behind these effects remained unclear and unexplained. In this study, we explore how vortex dynamics influence an independently actuated caudal fin in a carangiform swimmer.

Our present work investigates the hydrodynamic performance of a tandem pair of an undulating body with a passive independently pitching for a range of stiffness, damping, moment of Inertia, and Strouhal frequency.  Specifically, we employ a dynamic meshing fluid-structure interactions (FSI) based framework in \texttt{OpenFOAM}~v2312 to solve the Navier-Stokes equations at Reynolds numbers (Re)$=500$ and $5000$ and to capture the pitching response of the downstream foil, modeling a caudal fin. Our specific research objectives include: (i) quantification of how passive kinematics alter thrust generation of the tail compared to an active pitching of the tail, (ii) identification of which alteration in the dynamic characteristics of the body-tail joint improves the hydrodynamic performance of the swimmer, and (iii) elucidation of the the underlying vortex–body interactions that govern these performance trends. By addressing these important and unanswered questions until now about robotic fish-like platforms, our current work aims to contribute towards providing the guidlines for designs of next-generation bio-inspired AUVs.
%%%%%%%%%%%%%%%%%%%%%%%%%%%%%%%%%%%%%%%%%%%%%%%%%%%%%%%%%%%%%%%%%%%%%%%%%%%%%%%
\section{COMPUTATIONAL METHODOLOGY}

%%%%%%%%%%%%%%%%%%%%%%%%%%%%%%%%%%%%%%%%%%%%%%%%%%%%%%%%%%%%%%%%%%%%%%%%%%%%%%%
\subsection{Geometry and kinematics}

\begin{figure}[ht!]
    \centering
    \includegraphics[width=1.0\linewidth]{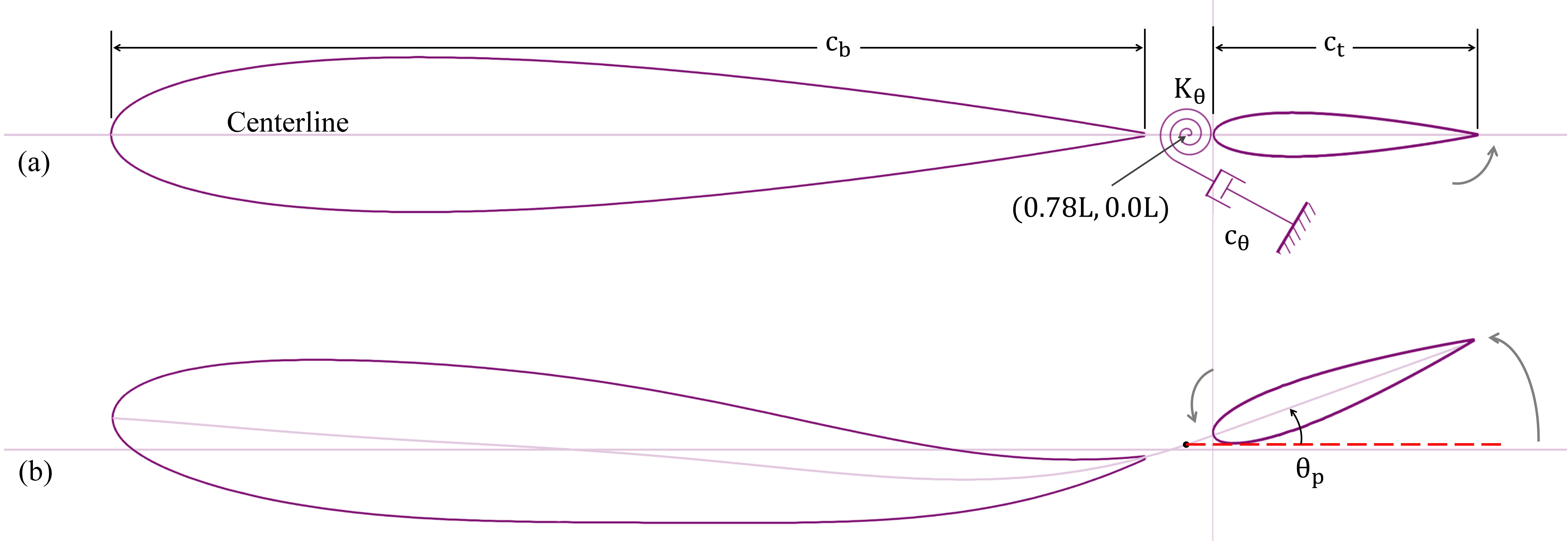}
    \caption{Two-dimensional model of the fish at (a) static position (${t}/{\tau} =0$), and (b) mid-oscillation instant (${t}/{\tau} =0.55$).}
    \label{fig:geometry}
\end{figure}

We model the body and tail of a carangiform-like swimmer by using two foils with an overall length of $L$. The body modeled using a {NACA-0015} foil is set to $c_b = 0.75L$, and length of the tail using a {NACA-0012} foil is set up with a length of $c_t = 0.20L$. There is a small gap between the main body and the tail which is set to $0.05L$, (see Fig.\ref{fig:geometry}), following the geometric features presented by by Gao et al, \cite{gao2018independent}. The body follows a carangiform undulating swimming flexure, with a continuous passive pitch control over the tail. The amplitude ($A\left(\frac{x}{L}\right)$) of the carangiform undulation is described by the following equation \cite{khalid2020flow, kamran2024does, khalid2021larger, khalid2016hydrodynamics}.

\begin{align}
A\left(\frac{x}{L}\right) &= 0.02 - 0.085\left(\frac{x}{L}\right) + 0.1625\left(\frac{x}{L}\right)^2; \,\,\,\, 0 < \frac{x}{L} < c_b
\label{eq:eq1}
\end{align}

The foil undulates with a frequency $f$, and the tail follows a continuous trajectory of the body by heaving at the undulation frequency, and tracing the amplitude ($h$) of the virtual joint at its peduncle, $h = A(0.78)$. The oscillating motion and the undulating motion is modeled by the mathematical expression, as presented as Eq. \ref{eq:eqmathModel}. The motion of the body and tail is initiated using a hyperbolic tangent function (also referred to as a Sigmoid function), given in Eq.~\ref{eq:hyper}, to ensure a smooth transition to the peak amplitude of undulation, active heaving, and active pitching. In this formulation, the exponent is cubed to produce a faster yet smoother rise to the peak amplitude. Notably, applying the Sigmoid function enables the passively pitching tail to reach steady-state oscillation much faster than without it, thereby reducing the overall computational cost.

\begin{align}
y\left(\frac{x}{L}\right) &= f\left(t\right)\,A\left(\frac{x}{L}\right) \cos\left[2\pi\left({\frac{x}{\lambda}}-{f}{t}\right)\right]
\label{eq:eqmathModel}
\end{align}

\begin{align}
f\left(t\right) &= \left[\frac{e^{3t} - 1}{e^{3t} + 1}\right]
\label{eq:hyper}
\end{align}

Figure. \ref{fig:geometry} shows the modeled flexure of the undulating body with a passively pitching tail. The tail is attached to a torsional spring and a viscous damper at the peduncle as illustrated in Fig. \ref{fig:geometry}a. As the undulation begins, the body and the tail follow the trajectory as shown in Fig. \ref{fig:geometry}b, where $\theta_p$ is the pitching angle that the tail makes with the centerline of the whole body when it is straight and stationary. The dashed line in Fig. \ref{fig:geometry}b shows the location of peduncle.

\noindent\hspace*{2em}The pitching dynamics of the tail is governed by the following equation:

\begin{equation}
    \left[\frac{2 \, c_{t}^{2}}{U_{\infty}^{2}} \, J^{*}\right] \ddot{\theta}_p+\left[\frac{2 \, c_{t}}{U_{\infty}} \, C^{*}\right] \dot{\theta}_p+\left[2 \, K^{*}\right] \theta_p=C_{M}
    \label{eq:eq3}
\end{equation}

\[
J^{*} = \frac{J_\theta}{\rho\,s\,c_t^{4}},\;
K^{*} = \frac{K_\theta}{\rho\,U_{\infty}^{2}\,s\,c_t^{2}},\;
C^{*} = \frac{c_\theta}{\rho\,U_{\infty}\,s\,c_t^{3}},\;
C_{M} = \frac{2M}{\rho\,U_{\infty}^{2}\,s\,c_t^{2}}.
\]

\noindent Where, $\rho$ denotes density of the fluid; $U_{\infty}$, the freestream velocity, $s$ the span of the tail in the lateral (out-of-plane) direction (chosen as unity in this study); $M$ the hydrodynamic moment about the pitching axis (pitching about the peduncle); $J_\theta$ the mass moment of inertia about the pitching axis. Besides, $c_\theta$ and $K_\theta$ represent the torsional damping coefficient and stiffness of the body-tail joint, respectively. The dimensionless stiffness coefficient $(K^*)$ is set as a dependent variable on dimensionless Inertia $(J^*)$ as explained below in Eq. \ref{eq:dimlessStiffnessStiff}.

\begin{equation}
    K^{*}= J^{*} \left[\frac{\pi\, n \, c_t \, f^*}{a}\right]^2 
    \label{eq:dimlessStiffnessStiff}
\end{equation}

\begin{equation}
    f^{*} = \frac{{2af} }{U_{\infty}}
    \label{eq:dimlessStiffnessFrequency}
\end{equation}

In Eqs. \ref{eq:dimlessStiffnessStiff}, and \ref{eq:dimlessStiffnessFrequency}, $a$ is the tail-beat amplitude of the body at $x/L = 0.75$, and $f$ is the frequency at which the body undulates. Here, n represents the $n^{th}$ multiple of the natural frequency which triggers the super-harmonic resonance of the system and contributes towards the dominating frequency, resulting in pitching of the tail at the same frequency as of the heaving motio (For this study, $n=10$) \cite{sridhar1975nonlinear, baek2001response}. Determined based on several test simulations performed with different values of $n$. In our current work $n=10$ exhibits the best resonance between the undulating body, and heaving and independently pitching tail. The dimensionless damping coefficient ($C^*$) is determined by the damping ratio ($\zeta$), and dimensionless stiffness coefficient ($K^*$), according to Eq. \ref{eq:damping} \cite{kim2018effect}.

\begin{equation}
    C^{*} = 2 \, \zeta\,\sqrt{J^{*} K^{*}}
    \label{eq:damping}
\end{equation}

%%%%%%%%%%%%%%%%%%%%%%%%%%%%%%%%%%%%%%%%%%%%%%%%%%%%%%%%%%%%%%%%%%%%%%%%%%%%%%%
\subsection{Governing Equations for Fluid Flows}

OpenFoam is an open source CFD solver providing a variety of numerical techniques to compute different terms in the governing equations for fluid flows. Here, we directly solve the unsteady Navier-Stokes equations for two-dimensional fluid flows around the swimmer, as explained earlier. We disregard the three-dimensional effects of the body and the fluid flows in this study. Although it is important to analyze all the flow features in a three-dimensional study, it is equally important to first observe these effects using a two-dimensional setup. With the plan to enforce multiple variables to study the behavior of the independently pitching tail, we conducted 150 independent simulations. The two-dimensional analysis provides a strong foundation for investigating relevant parameters, which can then be extended to three dimensions for a more in-depth analysis of the findings from the current work \cite{khalid2016hydrodynamics}.
with different numerical models and CFD tools We conduct $2D$ direct Numerical Simulations (DNS) of the swimmers, disregarding the effects of $3D$ flow for this study. The mathematical model for the fluid flow is based on the following non-dimensional forms of the continuity and incompressible Navier-Stokes equations \cite{kamran2024does, khalid2016hydrodynamics}.

\begin{equation}
    \frac{\partial u_{j}}{\partial x_{j}}=0
\label{eq:eqCont}
\end{equation}

\begin{equation}
    \frac{\partial u_{i}}{\partial t}+\frac{\partial}{\partial x_{j}}\left(u_{i} u_{j}\right)=-\frac{1}{\rho} \frac{\partial p}{\partial x_{i}}+\nu \frac{\partial^{2} u_{i}}{\partial x_{j} \partial x_{j}}
\label{eq:eqNavStoke}
\end{equation}

\noindent Where $i,j = {1,2} $, the $u_i$ are the Cartesian components of the flow velocity, $p$ is the pressure, and $\rho$ is the density of the fluid. The temporal term in the governing equations are discretized by using an implicit backward difference scheme. The $PIMPLE$ algorithm is used to couple the pressure and velocity field in an iterative manner over the moving mesh. This algorithm combines the Pressure-Implicit with Splitting of Operators ($PISO$) algorithm and Semi-Implicit Method for Pressure-Linked Equations ($SIMPLE$) algorithm. The convergence criterion for the iterative solution at each time step is set to $10^{-04}$. In this study, a Laplace equation with inverse-distance diffusivity is used for dynamic meshing \cite{jasak2009dynamic}.

\begin{figure}[ht!]
    \centering
    \includegraphics[width=1\linewidth]{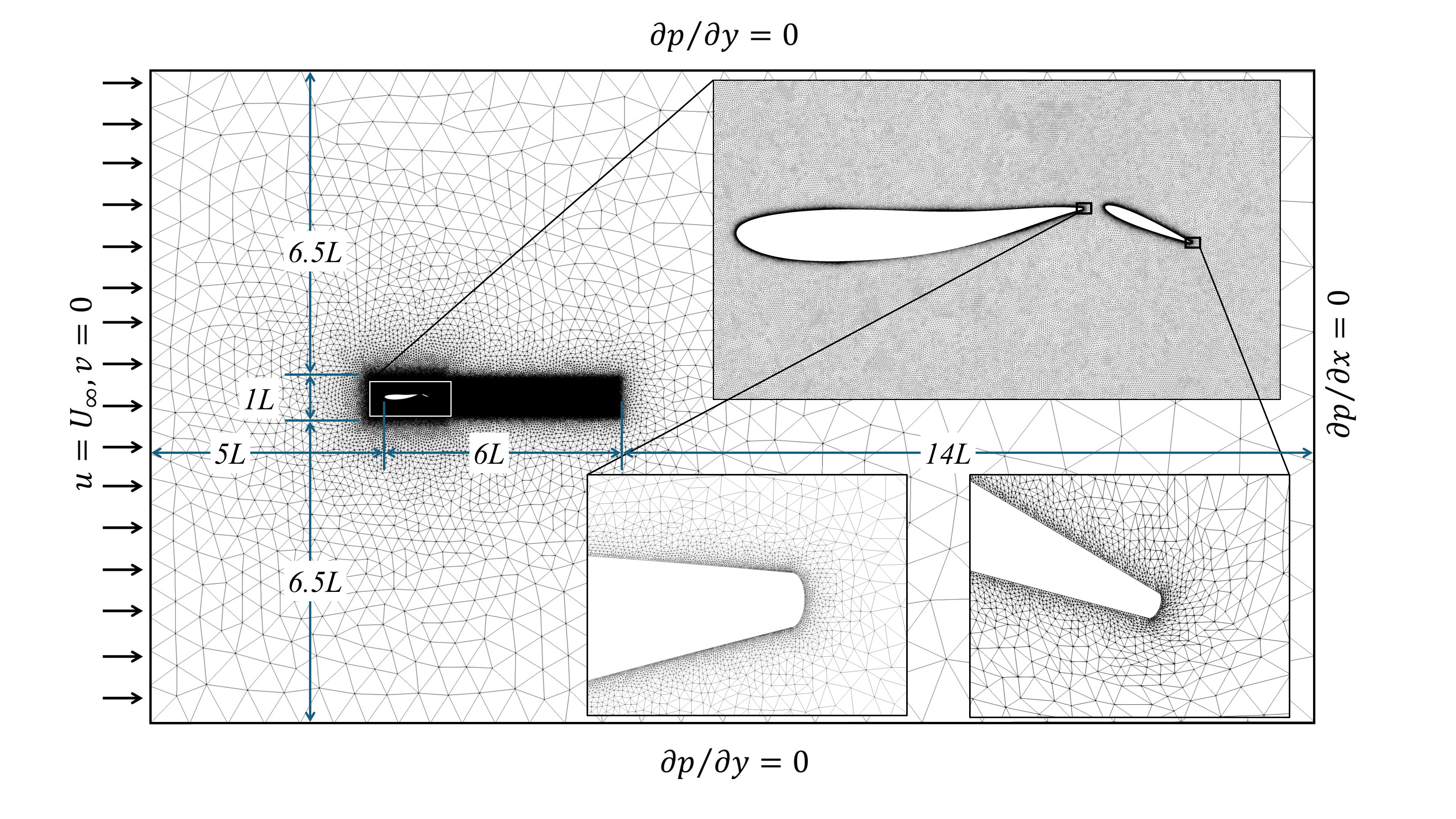}
    \caption{Flow domain and boundary conditions.}
    \label{fig:myMesh}
\end{figure}

An unstructured grid with a rectangular computational domain is used in this study. The grid illustrated in Fig. \ref{fig:myMesh} have boundaries selected to ensure minimal numerical errors. The velocity-inlet is at a distance of $5L$ from the leading edge of the body $LE_b$, and the pressure-outlet is set at a distance of $14L$ from the trailing edge of the tail ($TE_t$). In the downstream direction from $LE_b$ there is a refinement region of length $6L$ to capture the vortical flow features in the wake. The top and bottom boundaries are placed at a distance of $7L$ from the centerline. A uniform velocity field $U_\infty$ is set at the inlet, and the gauge pressure at the outlet, top and the bottom is set to zero (zero Gradient). OpenFoam is a Finite Volume Method ($FVM$) based solver which requires the $3D$ cells, but the front and back planes are defined as empty for $2D$ simulations(no interpolation in normal direction to the domain).

The computational grid deforms at every time step by solving a laplacian displacement equation, $\nabla \cdot (\gamma \nabla \xi) = 0$. An inverse-distance diffusivity ($\gamma=1/d$) referenced the body and tail enforces near-rigid motion close to the foils which smoothly dissipates away from foils. The dimensions of the grid are selected to also ensure that there are enough cells between the boundaries of the domain and the body itself to accommodate the moving mesh. 

%%%%%%%%%%%%%%%%%%%%%%%%%%%%%%%%%%%%%%%%%%%%%%%%%%%%%%%%%%%%%%%%%%%%%%%%%%%%%%%
\subsection{Verification \& Validation}

\begin{figure}[ht!]
    \centering
    \includegraphics[width=1\linewidth]{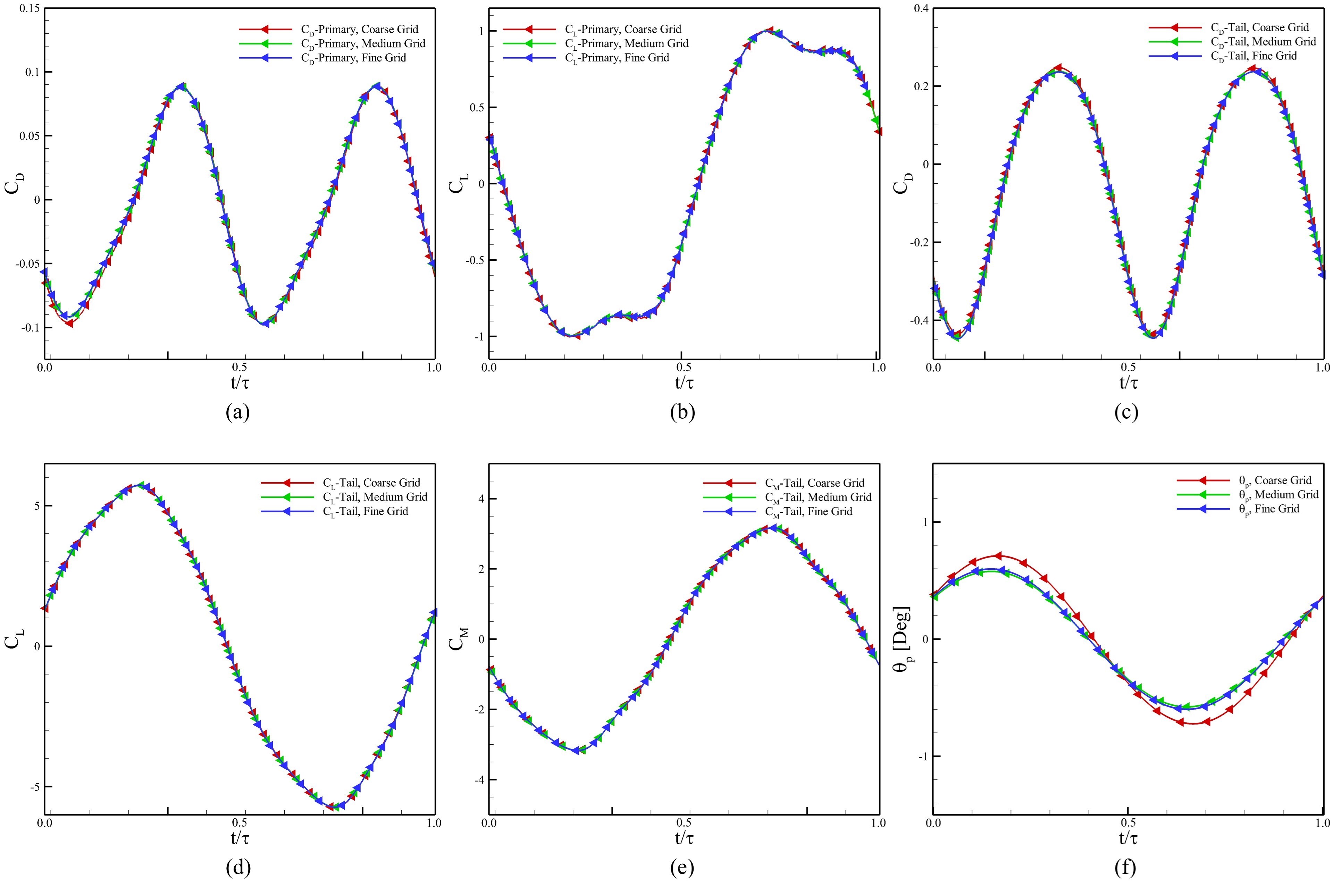}
    \caption{Results for convergence of Grid size.}
    \label{fig:gridInd}
\end{figure}

\begin{figure}[ht!]
    \centering  \includegraphics[width=1\linewidth]{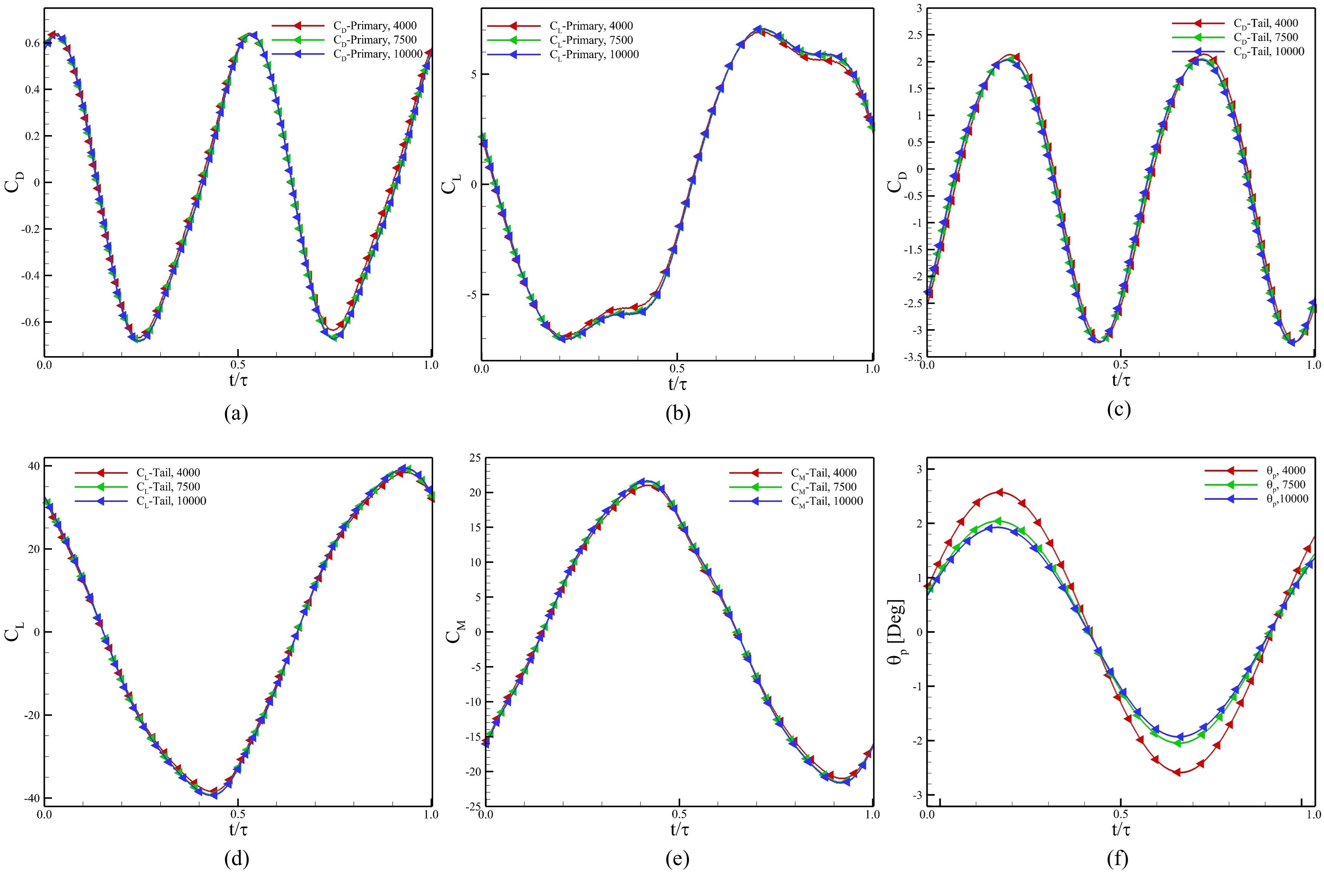}
    \caption{Results for convergence of Time step.}
    \label{fig:timeStep}
\end{figure}

We choose three grid resolutions to ensure the verification of our computational settings. For the grid-convergence and time step-independent studies, the Reynolds number is set as $500$, $J^*=0.125$, $f^*=0.6$, and $\zeta=1.0$. We select the final grid based on the reduction in relative error: Grid 1 with $307,489$ cells (Coarse Grid), Grid 2 with $483,623$ cells (Medium Grid), and Grid 3 with $651,687$ cells (Fine Grid). The body undergoes undulation, and the tail exhibits active heaving and passive pitching. The results of our simulations using the three grids are presented in Fig. \ref{fig:gridInd}a - \ref{fig:gridInd}f to compare the hydrodynamic force and moment coefficients of both the body and the tail as well as the pitching angle of the tail for the complete undulation cycle. Here $C_L$, $C_D$, and $C_M$ denote the lift coefficient, drag coefficient, and moment coefficient, respectively. Although there is no significant difference observed in the computed values of $C_L$, $C_D$, and $C_M$ for the body and tail, a clear improvement in capturing the pitching dynamics of the tail is found when the grid resolution is increased from the coarse to the medium grid. Based on these results, we select the medium grid for our next simulations. Next, we proceed for the time step-independence study, for which we choose three values of the time step size ($\Delta t$) so that the results stays unaffected by the selection of $\Delta t$. We choose $4000, 7500,$ and $10000$ time steps per oscillation cycles for this study. Figure \ref{fig:timeStep} shows the comparison of the hydrodynamic force and moment coefficients for the body and the tail for one complete undulation cycle. Based on the profiles of the pitching angle of the tail, we choose to proceed with $\Delta t$ according to $7500$ time steps in one undulation cycle.

\begin{figure}[ht!]
    \centering
    \includegraphics[width=1\linewidth]{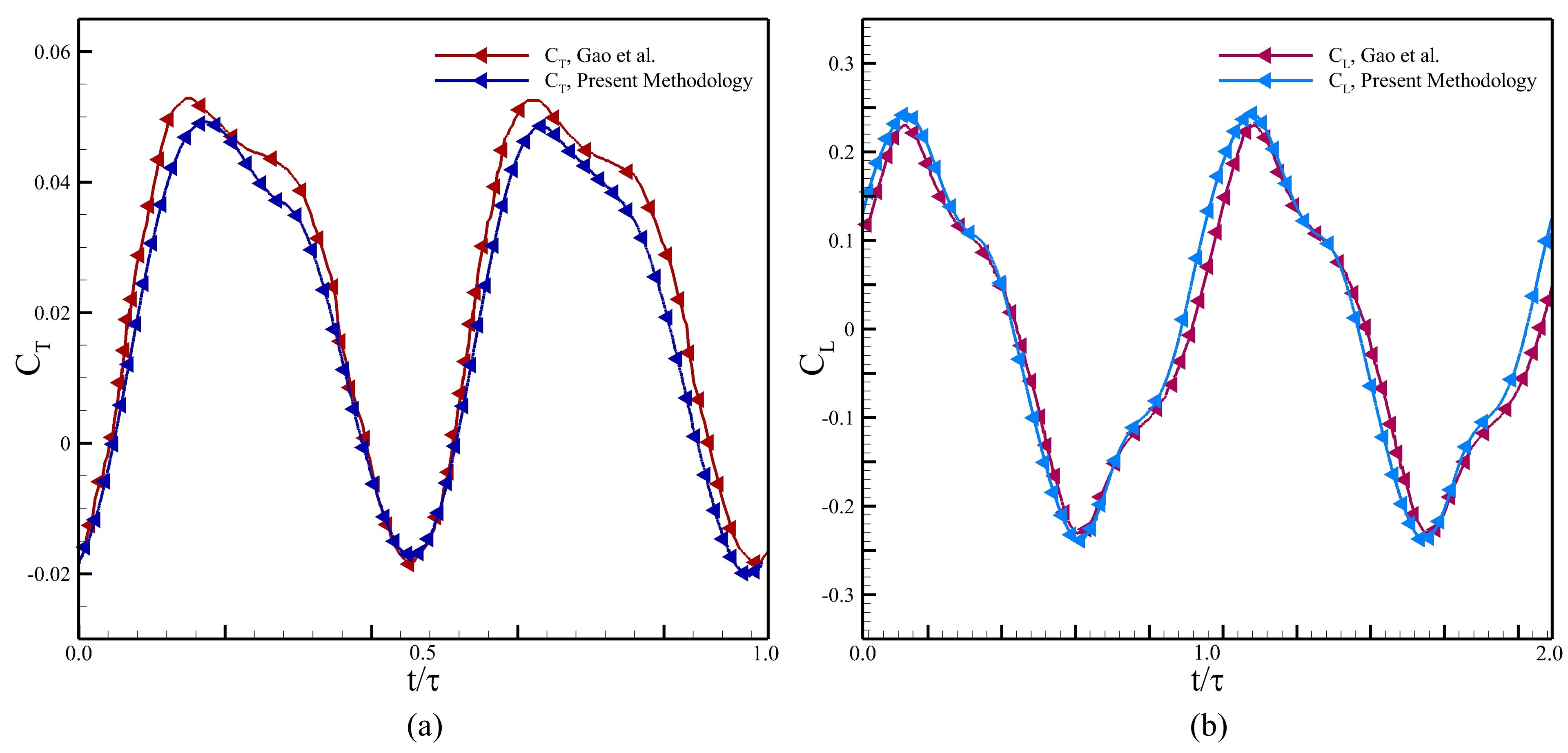}
    \caption{Comparison of our results (a) $C_T$, and (b) $C_L$ with those of Gao et al \cite{gao2018independent}.}
    \label{fig:validation}
\end{figure}

Our computational methodology and the strategy for the morphing mesh is validated by comparing the $C_T$, $C_L$ with Gao et al. \cite{gao2018independent}.Here, the Undulatory flexure over the main body $c_b$ is defined with a Gaussian envelope, whereas the kinematic parameters used for the validation study are defined in Table \ref{tab:validation}. The force coefficients, $C_T$ and $C_L$ are employed here, where the comparison of the present results with those rfom Gao et al. \cite{gao2018independent} is presented in Fig. \ref{fig:validation}. Please note that $C_T = -C_D$. We find a very close agreement between both sets of data, that indicates the accuracy of our simulation methodology.

\begin{table}[ht!]
  \centering
  \caption{Kinematic parameters used for the validation of our simulation methodology.}
  \label{tab:validation}
  \begin{tabular}{llc}
    \toprule
    Parameter & Symbol & Value \\ 
    \midrule
    Reynolds number                    & Re            & 5000 \\
    Ratio between the heave amplitude and the length of the body          & $h/c_t$           & 0.30 \\
    strouhal frequency                    & $f^*$            & 0.35 \\
    Phase between the heaving and pitching motion of the tail             & $\phi_{p}$      & 1.47 \\
    Undulating amplitude for the body        & $a_{0}$         & 0.045 \\
    \bottomrule
  \end{tabular}
\end{table}

%%%%%%%%%%%%%%%%%%%%%%%%%%%%%%%%%%%%%%%%%%%%%%%%%%%%%%%%%%%%%%%%%%%%%%%%%%%%%%%
\section{RESULTS \& DISCUSSION}

The kinematic parameters are selected to ensure effective simulation performance. The wavelength is fixed at $\lambda = 1.0$. The Strouhal frequency $f^*$ and damping ratio $\zeta$ are varied in increments of $0.1$ and $0.2$, respectively, within the ranges shown in Table~\ref{tab:discussion_table_1}. The non-dimensional inertia $J^*$ is adjusted in steps of $0.125$. The torsional spring stiffness for passive pitching depends on $J^*$, $n$, $c_t$, $a$, and $f^*$, as expressed in Equation~\ref{eq:dimlessStiffnessStiff}. Relevant kinematic parameters are summarized in Table~\ref{tab:discussion_table_1}.

\begin{table}[ht!]
  \centering
  \caption{Kinematic parameters.}
  \label{tab:discussion_table_1}
  \begin{tabular}{llc}
    \toprule
    Parameter & Symbol & Value \\
    \midrule
    Undulatory gait              & —               & Carangiform \\
    Strouhal frequency           & $f^*$            & $0.2$-$0.6$ \\
    Damping ratio                    & $\zeta$   & $0$-$1.0$ \\
    Dimensionless inertia & $J^{*}$   & $0.125$, $0.250$, $0.375$ \\
    Reynolds number              & Re            & $500$, $5000$ \\
    Primary-foil chord length    & $c_{b}$         & $0.75\,L$ \\
    Secondary-foil chord length  & $c_{t}$         & $0.20\,L$ \\
    \bottomrule
  \end{tabular}
\end{table}

\begin{figure}[ht!]
    \centering
    \includegraphics[width=1\linewidth]{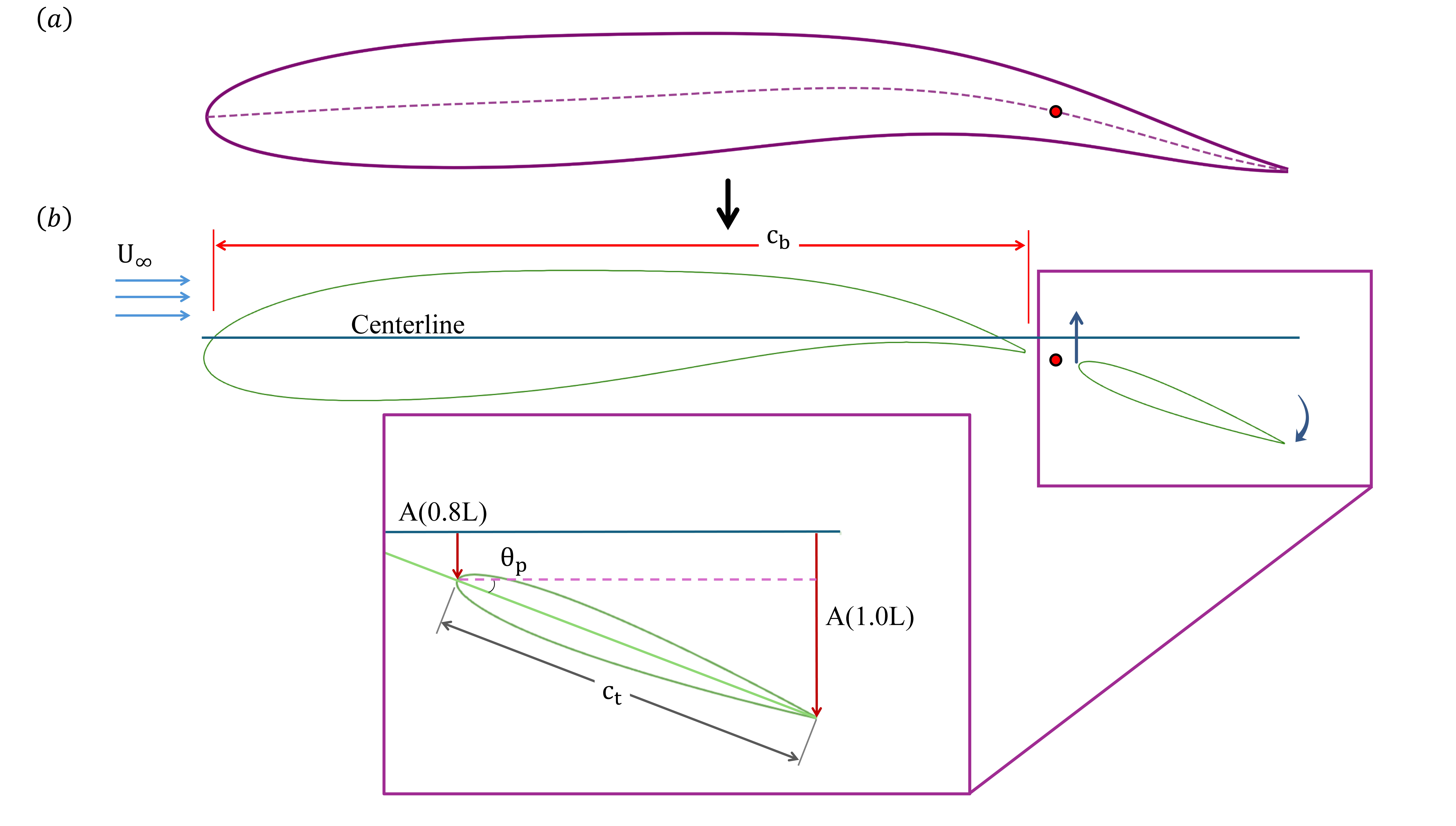}
    \caption{Model of the actively pitching tail mimicking a swimmer with continuous undulatory flexure.}
    \label{fig:c}
\end{figure}

A common modeling approach approximates the body of a carangiform swimmer as a single airfoil \cite{kamran2024does, khalid2016hydrodynamics, khalid2020flow, khalid2021larger, lucas2020airfoil}, as illustrated in Fig.~\ref{fig:c}a. In this study, we instead adopt a biomimetic two-foil configuration, representing the body and tail as separate foils, as shown in Fig.~\ref{fig:c}b. To ensure a consistent tail trajectory across all cases, the prescribed flapping motion is constrained so that the centerlines of the body and tail remain continuous. The amplitude trailing-edge amplitude of the tail is matched to the undulatory envelope of a carangiform swimmer at the posterior section of the body, with a maximum lateral amplitude of $A(1.0) = 0.1L$. Since the geometry and flow conditions are identical for both the actively and passively pitching tails, a direct comparison can be made to evaluate how hydrodynamic performance changes when the kinematics of the tail are switched from active to passive. In this framework, the heaving displacement of the tail is prescribed to follow the undulation of the body at its peduncle, while the kinematics of the tail with active pitching is defined by Eq.~\ref{eq:active}. 

\begin{equation}
\theta_{p}(t)=f\left(t\right)\,\sin^{-1}\!\Biggl\{
    \frac{1}{c_{t}}\Bigl[
        A(1.0)\,\cos\!\Bigl(2\pi\bigl(\tfrac{1.0}{\lambda}-ft\bigr)\Bigr) - A(0.8)\,\cos\!\Bigl(2\pi\bigl(\tfrac{0.8}{\lambda}-ft\bigr)\Bigr)
    \Bigr]
\Biggr\}.
\label{eq:active}
\end{equation}

\noindent where $theta _p$ denotes the angle.

For the actively pitching-tail configuration, a total of ten simulations are conducted at Re = $500$ and Re = $5000$, covering five Strouhal frequencies in the range $f^* = 0.2$–$0.6$. The maximum pitching angle of the tail, $(\theta_p)_{\max} = 30^\circ$, is defined as the angle between the leading edge and the centerline at its peak within an oscillation cycle (see Fig.~\ref{fig:c}). The resulting motion produces a phase difference ($\Psi$) of $112^\circ$ between the heaving and pitching of the tail. The performance of the tail is evaluated in terms of the ratio between the output power and input power, expressed as a power ratio ($\eta$) defined by Eq.~\ref{eq:power_ratio}. The power coefficients \(C_{P,\text{heave}}\), \(C_{P,\text{pitch}}\), and \(C_{P,\text{thrust}}\) are obtained from Eqs.~\eqref{eq:heavePower}–\eqref{eq:thrust}, following the methodology of Picard \emph{et~al.}\,\cite{picard2019oscillating} and Wu \emph{et~al.}\,\cite{wu2015numerical}. For the actively pitching cases, the \(C_{P,\text{thrust}}\) = \(C_{P,\text{out}}\).

\begin{equation}
\eta=  \frac{C_{P,\text{out}}}{C_{P,\text{in}}}
\label{eq:power_ratio}
\end{equation}

\begin{equation}
    C_{P,\text{heave}}(t) = C_{L}(t) \cdot \frac{\dot{h}_{\text{heave}}(t)}{U_{\infty}}
    \label{eq:heavePower}
\end{equation}

\begin{equation}
    C_{P,\text{pitch}}(t) = C_{M,\text{pitch}}(t) \cdot \frac{\dot{\theta}_{\text{pitch}}(t) \cdot c_t}{U_{\infty}}
    \label{eq:pitchPower}
\end{equation}

\begin{equation}
    C_{P,\text{thrust}}(t) = -C_{D}(t)
    \label{eq:thrust}
\end{equation}

\begin{table}[ht!]
  \centering
  \caption{Time-averaged power coefficients and power ratio ($\eta$) of the actively pitching tail at Re = $500$ and Re = $5000$.}
  \label{tab:activePerformance}
  %--- column setup --------------------------------------------------------
  % 1st col = f*, then five metrics for Re=500, then five for Re=5000
  \begin{tabular}{
      l % f*
      *{5}{S[table-format=2.3]} % 5 numeric cols for Re=500
      *{5}{S[table-format=2.3]} % 5 numeric cols for Re=5000
    }
    \toprule
    & \multicolumn{5}{c}{Re=$500$} & \multicolumn{5}{c}{Re=$5000$} \\ 
    \cmidrule(lr){2-6} \cmidrule(lr){7-11}
    {$f^{*}$} &
    {$\overline{C\!P}_{\text{heave}}$} &
    {$\overline{C\!P}_{\text{pitch}}$} &
    {$\overline{C\!P}_{\text{out}}$} &
    {$\overline{C\!P}_{\text{in}}$} &
    {$\eta\,[\%]$} &
    {$\overline{C\!P}_{\text{heave}}$} &
    {$\overline{C\!P}_{\text{pitch}}$} &
    {$\overline{C\!P}_{\text{out}}$} &
    {$\overline{C\!P}_{\text{in}}$} &
    {$\eta\,[\%]$} \\
    \midrule
    0.20 & 0.434 & 0.168 & 0.003 & 0.602 &  0.4 & 0.226 & 0.451 & 0.145 & 0.677 & 21.4 \\
    0.30 & 1.530 & 0.957 & 0.260 & 2.487 & 10.4 & 1.099 & 1.621 & 0.611 & 2.720 & 22.5 \\
    0.40 & 3.723 & 2.798 & 0.750 & 6.521 & 11.5 & 3.016 & 4.121 & 1.399 & 7.138 & 19.6 \\
    0.50 & 7.389 & 6.099 & 1.484 &13.489 & 11.0 & 6.462 & 8.338 & 2.514 &14.799 & 17.0 \\
    0.60 &12.903 &11.231 & 2.465 &24.134 & 10.2 &11.807 &14.767 & 3.959 &26.574 & 14.9 \\
    \bottomrule
  \end{tabular}
\end{table}

Here, $\dot{h}_{\text{heave}}$ and $\dot{\theta}_{\text{pitch}}$ denote the heaving and pitching velocities of the tail, respectively. The normalized mean power consumed by the tail to heave ($\overline{C}_{P,\text{heave}}$) and pitch ($\overline{C}_{P,\text{pitch}}$), alongside the output power ($\overline{C}_{P,\text{out}}$) for the actively pitching tail, are computed at Re = $500$ and Re = $5000$. As summarized in Table~\ref{tab:activePerformance}, the power-ratio $\eta$ increases significantly between Re =$500$ to $5000$, indicating more favorable energy transfer at Reynolds number $5000$. A consistent drop in $\eta$ is observed at $f^* = 0.6$ for both Reynolds numbers, while a low value of power ratio ($\eta = 0.4\%$) occurs at $Re = 500$ and $f^* = 0.2$, corresponding to negligible thrust output ($\overline{C}_{\text{T}} = -\overline{C}_{\text{D}} \approx 0$) as per the relationship from Eq.~\ref{eq:power_ratio}.

% ---------- FIGURE ----------
\begin{table}[ht!]
  \centering
  \caption{Time-averaged drag coefficients of the
           actively pitching tail at $Re = 500$ and $Re = 5000$.}
  \label{tab:activeThrust}
  %--- column setup --------------------------------------------------------
  % 1st col = f*, then two metrics for Re 500, then two for Re 5000
  \begin{tabular}{
      l % f*
      *{2}{S[table-format=2.3]} % numeric cols for Re 500
      *{2}{S[table-format=2.3]} % numeric cols for Re 5000
    }
    \toprule
    & \multicolumn{1}{c}{Re = 500} & \multicolumn{1}{c}{Re = 5000} \\ 
    \cmidrule(lr){2-3} \cmidrule(lr){4-5}
    {$f^{*}$} &
    {$\overline{C}_{\text{D, tail}}$} &
    {$\overline{C}_{\text{D, tail}}$} \\
    \midrule
    0.20 & -0.003 & -0.145  \\
    0.30 & -0.260 & -0.611 \\
    0.40 & -0.750 & -1.399 \\
    0.50 & -1.484 & -2.514 \\ 
    0.60 & -2.465 & -3.959\\
    \bottomrule
  \end{tabular}
\end{table}

Table~\ref{tab:activeThrust} presents the variation of the mean drag coefficient with an increasing $f^*$ for the tail of the swimmer. The results show a clear transition from near-zero thrust at low $f^*$ to increasingly negative values as the frequency rises, indicating a corresponding increase in thrust. Furthermore, at Re = $5000$, the negative mean drag coefficient (positive thrust) is consistently larger in magnitude than at Re=$500$, showing larger thrust generation at the higher Reynolds number.  
%%%%%%%%%%%%%%%%%%%%%%%%%%%%%%%%%%%%%%%%%%%%%%%%%%%%

%%%%%%%%%%%%%%%%%%   stability map   %%%%%%%%%%%%%%%%%

% ---------- FIGURE ----------
\begin{figure}[ht!]
    \centering
    \includegraphics[width=\linewidth]{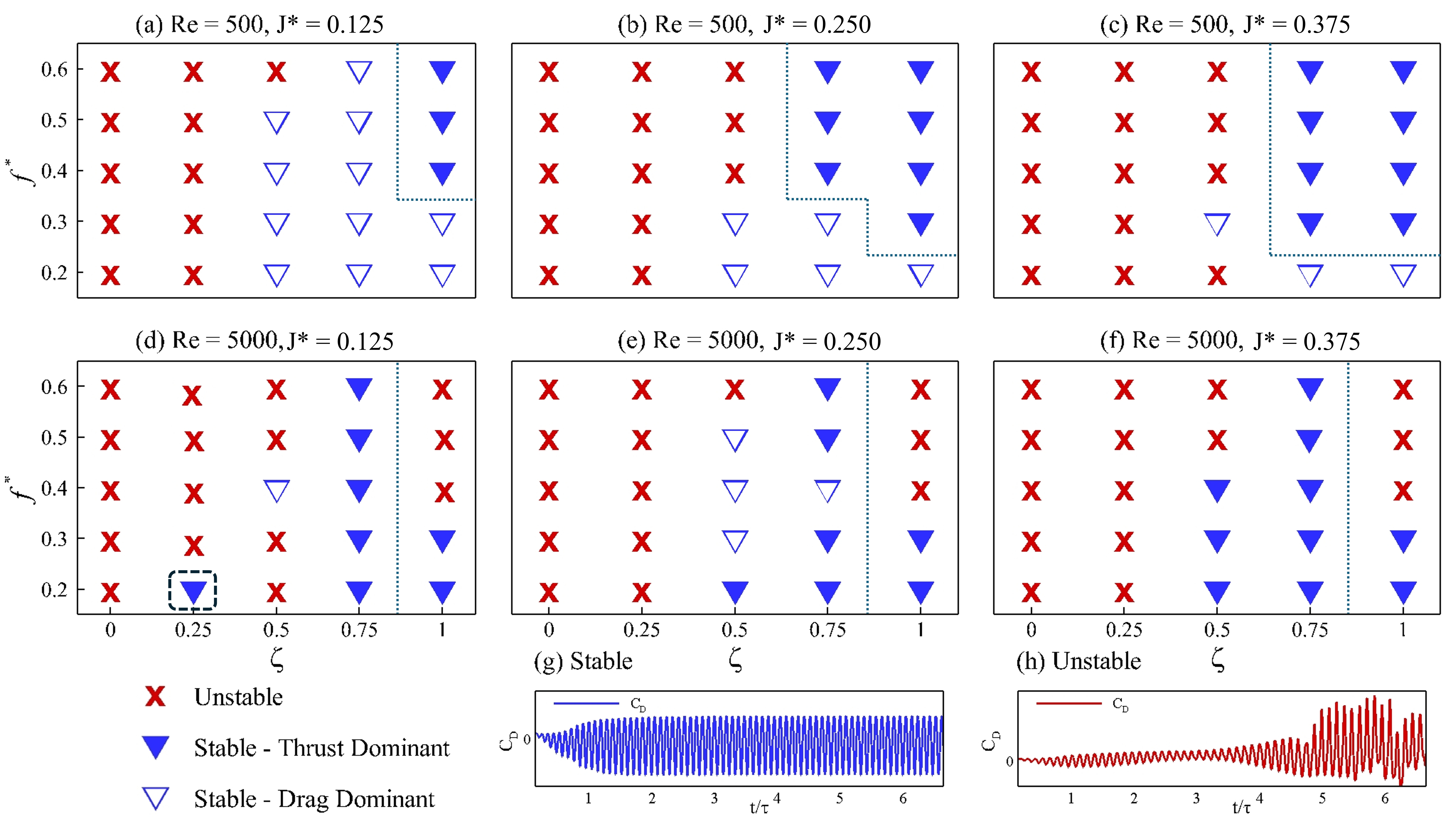}
    \caption{Stability map (a)-(f) insets illustrate $C_D$ for a typical stable (g) and unstable (h) cases.}
    \label{fig:map}
\end{figure}

Before discussing the dynamics of the swimmer with a passively pitching tail, we note that a total of $150$ numerical simulations are performed. The responses are categorized into three modes: stable drag-dominated, stable thrust-dominated, and unstable. This classification is summarized in Fig.~\ref{fig:map}, which we refer to as the stability maps, constructed for different sets of kinematic and flow parameters. The classification is based on $40$ oscillation cycles. A case is considered stable if the flow reaches a periodic steady state within this window; otherwise, it is classified as unstable. Stability is determined using the drag coefficient ($C_D$): if a case fails to reach a steady state (Fig.~\ref{fig:map}h), the case is labeled as unstable, whereas cases that converge to steady periodic behavior (Fig.~\ref{fig:map}g) are labeled stable. Although steady state is typically achieved within $5$–$8$ cycles, $40$ cycles are simulated to ensure complete decay of transient effects and convergence to steady state. Stable cases are further categorized using the mean drag coefficient ($\overline{C_D}$). If $\overline{C_D} > 0$, the case is classified as stable drag-dominated; if $\overline{C_D} < 0$, it is classified as stable thrust-dominated.

Across the stability maps, three clear patterns emerge. First, every undamped configuration ($\zeta=0$) diverges, confirming the requirement of minimal damping needed for a bounded motion of the passively pitching tail. Introducing a light damping ($\zeta = 0.25$) causes the tail to drift in and out of the thrust-producing region, highlighted by the dashed rectangle in Fig.~\ref{fig:map}(d), illustrating the decisive influence of fluid–structure coupling on the net force. At Re = $500$, stability improves with a higher inertia, with the tail exhibiting a stable dynamic response at Re = $500$, and the majority of the cases at Re = $5000$. In these cases, the response becomes thrust-dominated, as the $f^{*}$ increases from left to right. The consistent trends that we observe at Re = $500$ do not stay true when the Reynolds number is changed to $5000$. At Re = $5000$, the trend of stable and unstable cases stays consistent at $\zeta=0.75$ and $\zeta=1$, but the inconsistency based on the $f^*$ is prominent at $\zeta=0.25$ and $\zeta=0.50$

%%%%%%%%%%%%%%%%%%%%%%%%%%%%%%%%%%%%%%%%%%%%%%%%%%%%%%%%%%%%%%%%%%%%%%%%%%%%%%%

%%%%%%%%%%%%%%%%%%   mean drag   %%%%%%%%%%%%%%%%%

\begin{figure}[ht!]
    \centering
    \includegraphics[width=1\linewidth]{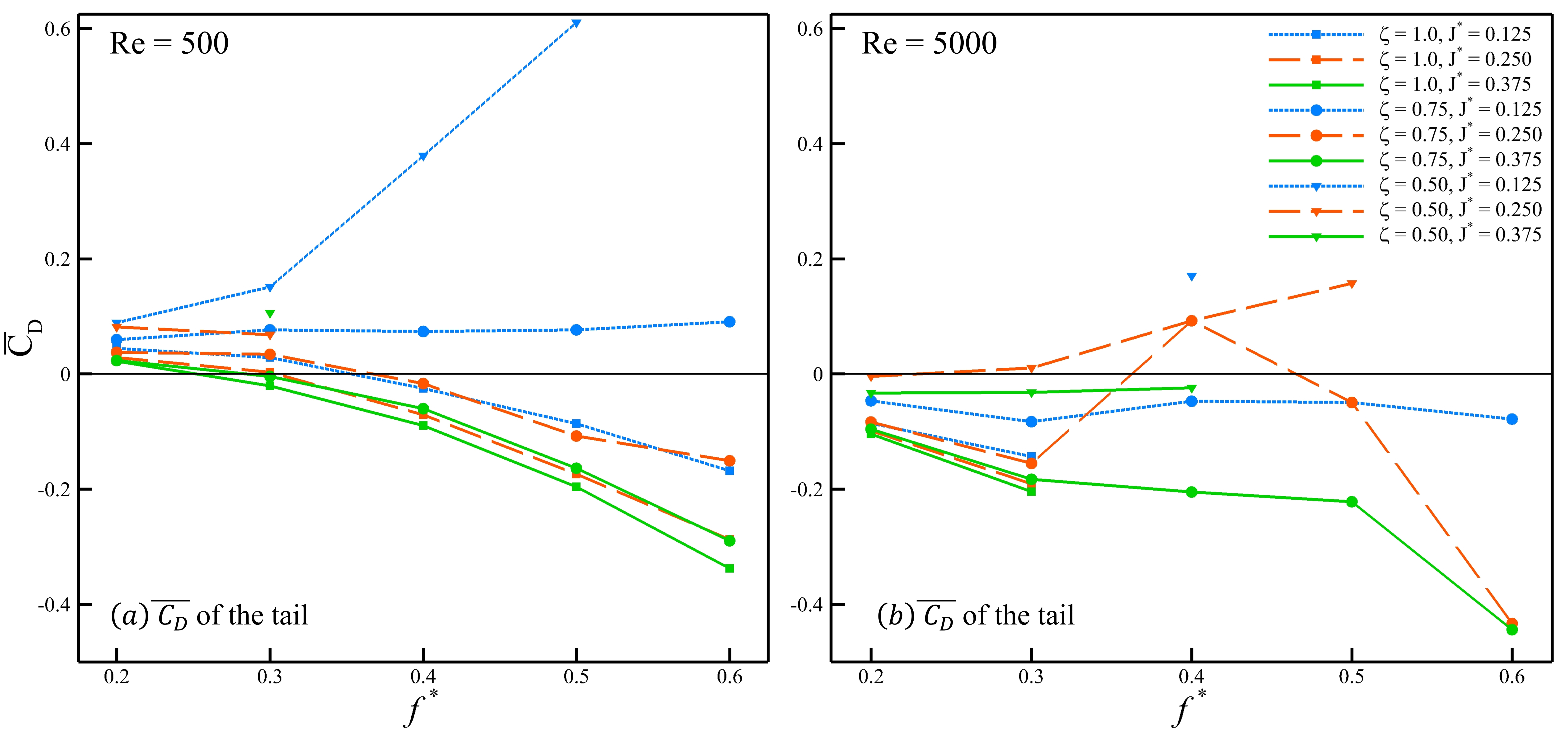}
    \caption{Mean drag coefficient of the passively pitching tail at (a) Re = $500$, and (b) $5000$.}
    \label{fig:thrustPlot}
\end{figure}

We now turn to the passively pitching tail and analyze its drag coefficient for all stable cases. It allows us to assess how the transition from active to passive pitching influences the performance of the swimmer, based on the mean drag coefficient presented in Fig.~\ref{fig:thrustPlot}. Figures~\ref{fig:thrustPlot}a and \ref{fig:thrustPlot}b show the mean drag coefficient ($\overline{C_D}$) of the tail at Re = $500$ and Re = $5000$, respectively. From Fig.~\ref{fig:thrustPlot}a, the first observation is that the configuration with $\zeta = 0.50$ and $J^* = 0.125$ exhibits an increasing $\overline{C_D}$ (i.e., decreasing thrust) as the $f^*$ increases to $0.5$. Beyond this point, at $f^* = 0.6$, the case becomes unstable.  Secondly, for the same value of inertia ($J^* = 0.125$) but with a higher $\zeta = 0.75$, the trend remains nearly constant across the range of Strouhal frequencies considered here. When this trend is compared to that of at $\zeta = 0.5$ a large shift, increasing the damping ratio shows a reduction of $\overline{C_D}$. Further increasing the damping ratio ($\zeta = 1.0$), the trend shows a behavior similar to that of the actively pitching tail where the thrust increases as the Strouhal frequency is increased. Finally, examining the stable cases with a passively pitching tail, at $\zeta = 0.50$ for a range of inertia values ($J^* = 0.125$–$0.375$), neither of the cases fully transition towards producing positive thrust ($\overline{C_D} < 0$). However, all of the cases display a reduction in drag with an increase in inertia, and this trend of drag reduction with a higher inertia is consistently observed for all damping ratios chosen for this work.

At Re = $5000$ shown in Fig.~\ref{fig:thrustPlot}b, the number of stable cases decreases, making it more difficult to identify clear trends and assess the effect of Strouhal frequency on the mean drag coefficient. Compared to the cases at Re = $500$, the configurations that remain stable at Re = $5000$ generally show better performance in terms of mean thrust. However, despite the improvement when the Reynolds number increases, the inconsistencies in the trends make it difficult to draw firm conclusions about the role of the Strouhal frequency. Based on the trends observed in Figs.~\ref{fig:thrustPlot}a and \ref{fig:thrustPlot}b, the behaviour of the swimmer about its peduncle can be interpreted by categorizing the tail into three types according to its inertia $J^*=0.125-0.375$. While the inertia defines the characteristic response of the tail itself, the damping ratio characterizes the nature of its joint to the body. A lower damping ratio corresponds to a more loosely joined and flexible body and tail connection, whereas a higher damping ratio indicates a stiffer and less flexible joint.

%%%%%%%%%%%%%%%%%%%%%%%%%%%%%%%%%%%%%%%%%%%%%%%%%%%%%%%%%%%%%%%%%%%%%%%%%
%%%%%%%%%%%%%%%%%%   Max-pitching angle plot   
%%%%%%%%%%%%%%%%%%%%%%%%%%%%%%%%%%%%%%%%%%%%%%%%%%%%%%%%%%%%%%%%%%%%%%%%%%%%%%%%%%%%%%%%%%%%%%%%%%%%%%%%%%%%%%%%%
\begin{table}[H]
  \centering
  \renewcommand{\arraystretch}{0.85}

  \begin{tabular}{cc}
  % ----------- Row 1: f* = 0.20 and 0.30 -----------------
  \multicolumn{1}{c}{\textbf{(a) $f^* = 0.20$}} &
  \multicolumn{1}{c}{\textbf{(b) $f^* = 0.30$}} \\
  \begin{tabular}{l l 
                  S[table-format=1.3] S[table-format=2.2]
                  S[table-format=1.3] S[table-format=2.2]}
    \toprule
    {$J^*$} & {$\zeta$} &
    \multicolumn{2}{c}{Re = $500$} &
    \multicolumn{2}{c}{Re = $5000$} \\
    \cmidrule(lr){3-4}\cmidrule(lr){5-6}
    & & {$\overline{C}_D$} & {$\theta_{p,\max}$} & {$\overline{C}_D$} & {$\theta_{p,\max}$} \\
    \midrule
    \multirow{3}{*}{0.125} & 0.50 & {0.09} & {7.18} & {N/A} & {N/A} \\
                           & 0.75 & {0.06} & {4.59} & {-0.05} & {4.84} \\
                           & 1.00 & {0.05} & {3.09} & {-0.09} & {3.40} \\
    \midrule
    \multirow{3}{*}{0.250} & 0.50 & {0.09} & {5.69} & {-0.01} & {6.84} \\
                           & 0.75 & {0.04} & {3.00} & {-0.08} & {3.07} \\
                           & 1.00 & {0.03} & {1.93} & {-0.10} & {1.94} \\
    \midrule
    \multirow{3}{*}{0.375} & 0.50 & {N/A}    & {N/A}    & {-0.03} & {5.68} \\
                           & 0.75 & {0.02} & {1.34} & {-0.10} & {2.21} \\
                           & 1.00 & {0.02} & {1.34} & {-0.10} & {1.34} \\
    \bottomrule
  \end{tabular}
  &
  \begin{tabular}{l l 
                  S[table-format=1.3] S[table-format=2.2]
                  S[table-format=1.3] S[table-format=2.2]}
    \toprule
    {$J^*$} & {$\zeta$} &
    \multicolumn{2}{c}{Re = $500$} &
    \multicolumn{2}{c}{Re = $5000$} \\
    \cmidrule(lr){3-4}\cmidrule(lr){5-6}
    & & {$\overline{C}_D$} & {$\theta_{p,\max}$} & {$\overline{C}_D$} & {$\theta_{p,\max}$} \\
    \midrule
    \multirow{3}{*}{0.125} & 0.50 & {0.15} & {7.26} & {N/A} & {N/A} \\
                           & 0.75 & {0.08} & {4.62} & {-0.08} & {4.53} \\
                           & 1.00 & {0.03} & {3.32} & {-0.14} & {3.18} \\
    \midrule
    \multirow{3}{*}{0.250} & 0.50 & {0.07} & {4.68} & {0.01} & {6.41} \\
                           & 0.75 & {0.04} & {3.01} & {-0.16} & {2.88} \\
                           & 1.00 & {0.01} & {1.83} & {-0.19} & {1.84} \\
    \midrule
    \multirow{3}{*}{0.375} & 0.50 & {0.11}    & {5.24}    & {-0.03} & {5.16} \\
                           & 0.75 & {-0.01} & {2.20} & {-0.18} & {2.09} \\
                           & 1.00 & {-0.02} & {1.35} & {-0.20} & {1.28} \\
    \bottomrule
  \end{tabular}
  \\[1.0em]

  % ----------- Row 2: f* = 0.40 and 0.50 -----------------
  \multicolumn{1}{c}{\textbf{(c) $f^* = 0.40$}} &
  \multicolumn{1}{c}{\textbf{(d) $f^* = 0.50$}} \\
  \begin{tabular}{l l S[table-format=1.3] S[table-format=2.2]
                      S[table-format=1.3] S[table-format=2.2]}
    \toprule
    {$J^*$} & {$\zeta$} &
    \multicolumn{2}{c}{Re = $500$} &
    \multicolumn{2}{c}{Re = $5000$} \\
    \cmidrule(lr){3-4}\cmidrule(lr){5-6}
    & & {$\overline{C}_D$} & {$\theta_{p,\max}$} & {$\overline{C}_D$} & {$\theta_{p,\max}$} \\
    \midrule
    \multirow{3}{*}{0.125} & 0.50 & {0.38} & {8.16} & {0.17} & {6.30} \\
                       & 0.75 & {0.07} & {5.12} & {-0.05} & {4.30} \\
                       & 1.00 & {-0.03} & {3.54} & {N/A} & {N/A} \\
    \midrule
    \multirow{3}{*}{0.250} & 0.50 & {N/A} & {N/A} & {0.09} & {5.89} \\
                       & 0.75 & {-0.02} & {3.19} & {0.09} & {5.89} \\
                       & 1.00 & {-0.07} & {1.98} & {N/A} & {N/A} \\
    \midrule
    \multirow{3}{*}{0.375} & 0.50 & {N/A} & {N/A} & {-0.02} & {4.99} \\
                       & 0.75 & {-0.06} & {2.21} & {-0.21} & {2.12} \\
                       & 1.00 & {-0.09} & {1.35} & {N/A} & {N/A} \\
    \bottomrule
  \end{tabular}
  &
  \begin{tabular}{l l S[table-format=1.3] S[table-format=2.2]
                      S[table-format=1.3] S[table-format=2.2]}
    \toprule
    {$J^*$} & {$\zeta$} &
    \multicolumn{2}{c}{Re = $500$} &
    \multicolumn{2}{c}{Re = $5000$} \\
    \cmidrule(lr){3-4}\cmidrule(lr){5-6}
    & & {$\overline{C}_D$} & {$\theta_{p,\max}$} & {$\overline{C}_D$} & {$\theta_{p,\max}$} \\
    \midrule
    \multirow{3}{*}{0.125} & 0.50 & {0.61} & {8.19} & {N/A} & {N/A} \\
                       & 0.75 & {0.08} & {5.12} & {-0.05} & {2.96} \\
                       & 1.00 & {-0.09} & {3.54} & {N/A} & {N/A} \\
    \midrule
    \multirow{3}{*}{0.250} & 0.50 & {N/A} & {N/A} & {0.16} & {6.17} \\
                       & 0.75 & {-0.11} & {3.18} & {-0.05} & {2.39} \\
                       & 1.00 & {-0.18} & {1.97} & {N/A} & {N/A} \\
    \midrule
    \multirow{3}{*}{0.375} & 0.50 & {N/A} & {N/A} & {N/A} & {N/A} \\
                       & 0.75 & {-0.17} & {2.28} & {N/A} & {N/A} \\
                       & 1.00 & {-0.20} & {1.37} & {N/A} & {N/A} \\
    \bottomrule
  \end{tabular}
  \\[1.0em]

  % ----------- Row 3: f* = 0.60 -----------------
  \multicolumn{2}{c}{\textbf{(e) $f^* = 0.60$}} \\
  \multicolumn{2}{c}{
    \begin{tabular}{l l S[table-format=1.3] S[table-format=2.2]
                        S[table-format=1.3] S[table-format=2.2]}
      \toprule
      {$J^*$} & {$\zeta$} &
      \multicolumn{2}{c}{Re = $500$} &
      \multicolumn{2}{c}{Re = $5000$} \\
      \cmidrule(lr){3-4}\cmidrule(lr){5-6}
      & & {$\overline{C}_D$} & {$\theta_{p,\max}$} & {$\overline{C}_D$} & {$\theta_{p,\max}$} \\
    \midrule
    \multirow{3}{*}{0.125} & 0.50 & {N/A} & {N/A} & {N/A} & {N/A} \\
                       & 0.75 & {0.09} & {5.23} & {-0.08} & {4.12} \\
                       & 1.00 & {-0.17} & {3.63} & {N/A} & {N/A} \\
    \midrule
    \multirow{3}{*}{0.250} & 0.50 & {N/A} & {N/A} & {N/A} & {N/A} \\
                       & 0.75 & {-0.15} & {3.24} & {-0.44} & {2.98} \\
                       & 1.00 & {-0.29} & {2.02} & {N/A} & {N/A} \\
    \midrule
    \multirow{3}{*}{0.375} & 0.50 & {N/A} & {N/A} & {N/A} & {N/A} \\
                       & 0.75 & {-0.30} & {2.31} & {-0.45} & {2.13} \\
                       & 1.00 & {-0.34} & {1.39} & {N/A} & {N/A} \\
    \bottomrule
    \end{tabular}
  } \\
  \end{tabular}
    \caption{Data on stable cases presenting the mean drag coefficient 
($\overline{C_D}$) and maximum pitching angle ($\theta_{p,\text{max}}$) 
of the passively pitching tail across Strouhal frequencies  
($f^*$), damping ratios ($\zeta$), and inertia ratios ($J^*$), at Reynolds numbers Re = $500$ and Re = $5000$.} 
  \label{tab:stableAll}
\end{table}

The nature of the joint between the body and the tail of the swimmer influences its thrust performance, as discussed earlier. From the previous analysis, a stiffer joint between the body and the tail corresponds to higher thrust or a more thrust-dominant behavior, whereas a more flexible or looser joint leads to a drag-dominant response. Table~\ref{tab:stableAll} presents the maximum pitching angle ($\theta_{p,\text{max}}$) alongside the mean drag coefficient ($\overline{C_D}$) of the tail at Re = $500$ and Re = $5000$ for all stable cases. The results are grouped in Table~\ref{tab:stableAll}a--\ref{tab:stableAll}e according to the Strouhal frequency ($f^* = 0.2$--$0.6$). Here, \textit{N/A} indicates values that could not be evaluated due to the unstable nature of the configuration, as identified in the stability maps shown in Figs.~\ref{fig:map}a--\ref{fig:map}f. 

Analyzing the effect of Strouhal frequency, for a given configuration of $J^*$ and $\zeta$, an increase in $f^*$ leads to a larger maximum pitching angle ($\theta_{p,\text{max}}$). It is accompanied by a clear trend of decreasing mean drag coefficient ($\overline{C_D}$) or equivalently increasing thrust ($-\overline{C_D}$), a behavior consistent at both Re = $500$ and Re = $5000$. However, when $f^*$ and $\zeta$ are held constant to isolate the effect of $J^*$, $\theta_{p,\text{max}}$ decreases alongside decreasing $\overline{C_D}$. A similar trend is observed when $f^*$ and $J^*$ are fixed and  $\zeta$ is varied. These observations suggest that a larger pitching amplitude is beneficial only when achieved by increasing the Strouhal frequency. In contrast, when the larger amplitude results from reduced inertia or damping ratio at a fixed Strouhal frequency, it adversely affects thrust generation.
%%%%%%%%%%%%%%%%%%   Power Ratio   %%%%%%%%%%%%%%%%%

\begin{figure}[ht!]
    \centering
    \includegraphics[width=1\linewidth]{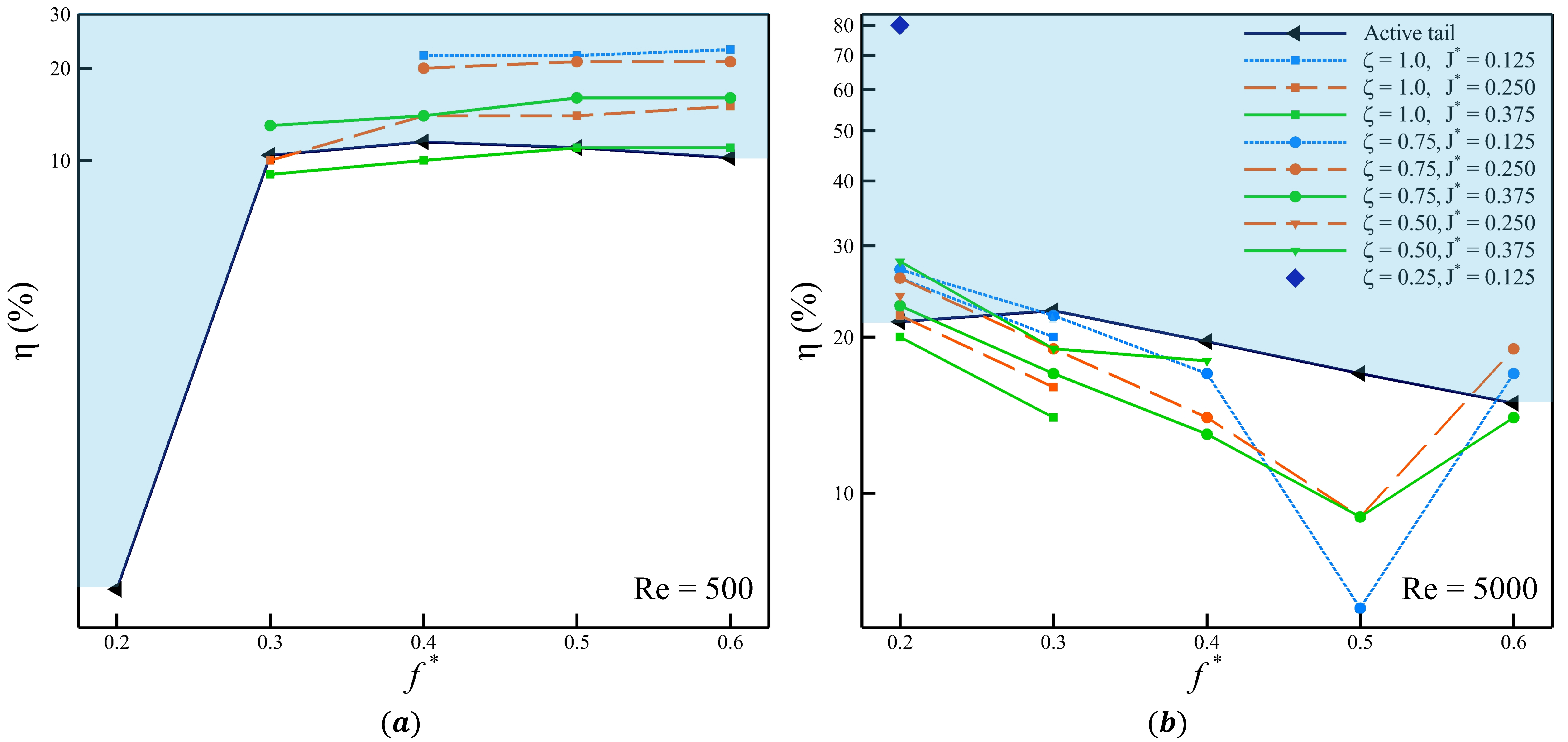}
    \caption{Power ratio at (a) Re = $500$, and (b) $5000$.}
    \label{fig:PowerRatioPlot}
\end{figure}

The swimmer with an actively pitching tail generates higher thrust than its passive counterparts. In these cases, a pitching amplitude of $30^\circ$ is required to ensure that the trailing edge of the tail spans the peak-to-peak amplitude of the carangiform swimmer ($A(L) = 0.1L$). The heaving amplitude of the tail, defined as the displacement of the peduncle, is $A(0.78) = 0.0525L$, which is relatively smaller than the pitching amplitude of an actively pitching tail. When such a large angular amplitude (pitching) is combined with a small linear amplitude (heaving), the power expenditure increases significantly, particularly at higher Strouhal frequencies. The larger peak-to-peak amplitude corresponds to greater thrust production. In contrast, for the passively pitching tail, the maximum observed pitching angle does not exceed $\theta_{p,\max} = 8.2^\circ$. This reduced pitching angle explains the lower thrust coefficients observed. However, it also results in a lower-power swimming mode, as the pitching of the tail requires less input power. These two contrasting behaviors are captured quantitatively by the power ratio ($\eta$), shown in Fig.~\ref{fig:PowerRatioPlot}. 

Figures~\ref{fig:PowerRatioPlot}a and \ref{fig:PowerRatioPlot}b show the $\eta$ as a function of $f^*$ for Re = $500$ and Re = $5000$, respectively. The power ratio is plotted on the logarithmic scale to capture the full range of values and trends. The solid black line represents the power ratio for the actively pitching tail, which serves as a reference for comparison with the cases with a passively pitching tail. The region above this line highlighted in blue represents the configurations where the tail with passive pitching outperforms the active counterpart. At Re = $500$, a significant number of cases with a passively pitching tail achieve higher $\eta$ than those with the active tail. As $f^*$ increases, the power ratio of the cases with a passive tail shows a modest upward trend, whereas the cases with the active tail only peaks at $f^* = 0.4$ and then decreases subsequently. Indicating that at lower Reynolds numbers, the passive tail often provides a performance advantage based on the $\eta$.

In contrast, at Re = $5000$, most of the cases with a passively pitching tail exhibit lower power ratios than their active counterparts. Only a limited number of configurations outperform the active case, and these are again concentrated near $f^* = 0.2$. Thus, the overall trend shifts with Reynolds number: passively pitching tails show superior performance at Re = $500$, while actively pitching tails dominate at Re = $5000$. When the passively pitching tail configurations are compared in terms of its thrust production and power ratio as performance metrics, a consistent trade-off emerges. Configurations that generate larger mean thrust tend to exhibit lower power ratios, whereas those with lower mean thrust generally achieve higher power ratios. An intermittent case at $Re = 5000$, with $\zeta = 0.25$ and $J^* = 0.125$, shows a unique transition from thrust-dominant to drag-dominant behaviour and back to thrust-dominant behaviour. This case reaches a maximum pitching angle of $25^\circ$ and is the only configuration among the 150 simulations to exhibit such dynamics. Given its rarity, This case is not examined further in the present study, primarily because no conclusive correlation or influence of any controlling parameter can be established.

From the quantitative analysis of the results, we observe a consistent trend for both the actively and passively pitching tails: as the Strouhal frequency increases, the thrust coefficient increases. To further investigate this phenomenon, we examine the vortex street in the wakes of the swimmers at Re = $500$, and $5000$. To keep the discussion centred on the role of Strouhal frequency, we consider all cases of the actively pitching tail for increasing $f^*$. For the passively pitching tail, we focus on configurations with fixed parameters $\zeta = 1.0$ and $J^* = 0.375$, varying only the Strouhal frequency. Both stable and unstable configurations with a passively pitching tails are included to provide additional insight into the origin of the instabilities observed at a higher $f^*$ seen earlier in the stability maps (Fig.~\ref{fig:map}). 

\begin{figure}[ht!]
    \centering
    \includegraphics[width=1\linewidth]{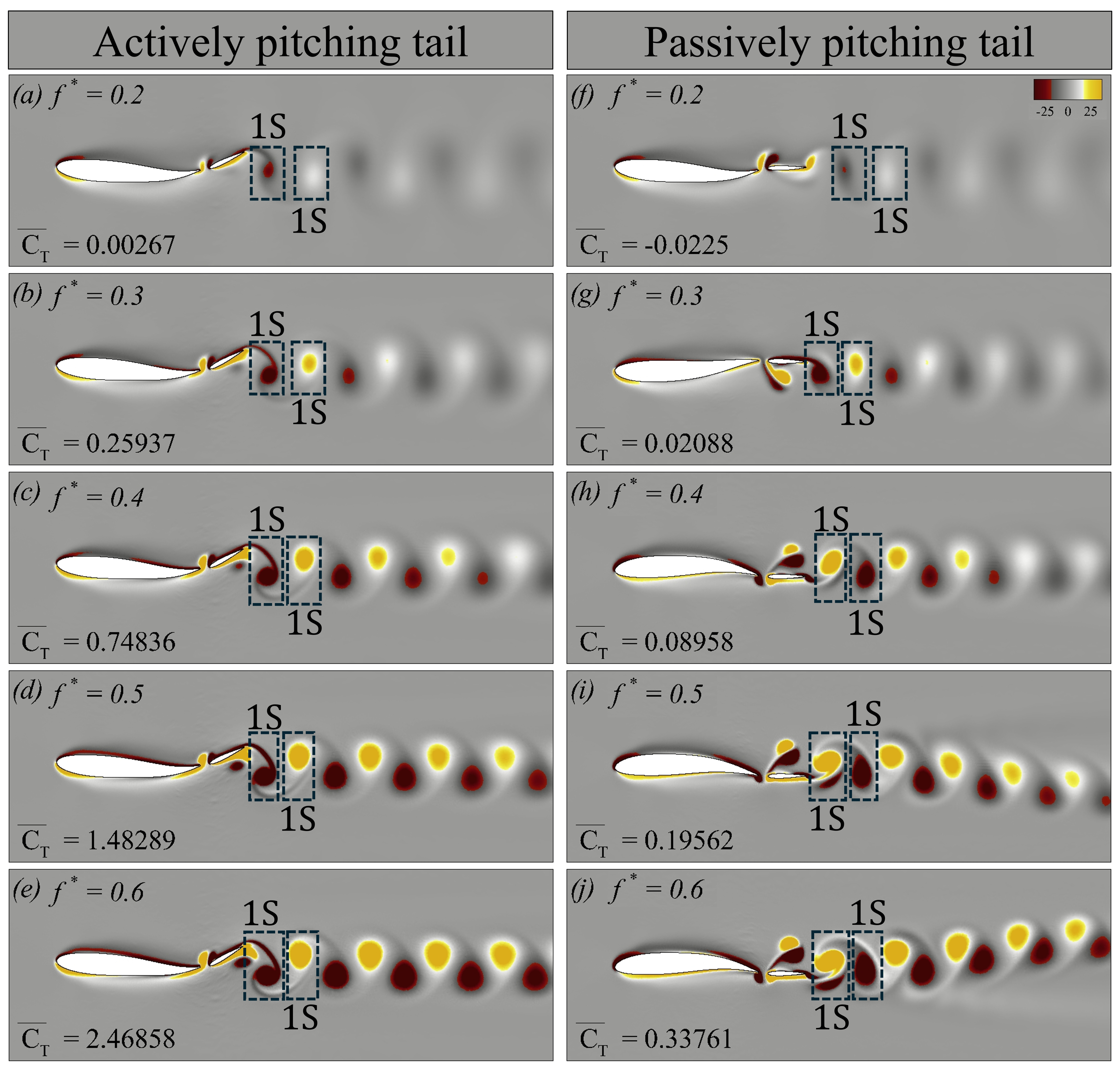}
    \caption{Vorticity contours corresponding to the swimmer with (a-e) an actively pitching tail, and (f-j) a passively pitching tail respectively at Re = $500$.}
    \label{fig:vorticity_comparison500}
\end{figure}

We first examine the vortex dynamics of the swimmers at Re = $500$, as shown in Fig.~\ref{fig:vorticity_comparison500}. The first column (Figs.~\ref{fig:vorticity_comparison500}a–\ref{fig:vorticity_comparison500}e) presents the contours of vorticity for cases with an actively pitching tail, whereas the second column (Figs.~\ref{fig:vorticity_comparison500}f–\ref{fig:vorticity_comparison500}j) corresponds to the cases with a passively pitching tail. For each case, we report report the $\overline{C_T}$, to quantify the propulsive performance associated with the observed wake structures. In Fig.~\ref{fig:vorticity_comparison500}, $1S$ denotes a single vortex shed into the wake during a half oscillation cycle.

From Figs.~\ref{fig:vorticity_comparison500}a and \ref{fig:vorticity_comparison500}f, we observe that at $f^* = 0.2$, the vortex street in the wake of both swimmers is faint. Nevertheless, even from this weak signatures, it is clear that both wakes form a classical von K\'arm\'an vortex street, typically associated with drag-dominated flows. For the actively pitching tail, although the mean thrust coefficient $\overline{C_T}$ is positive, its value is too small to generate an effective thrust. As the $f^*$ increases to $0.3$ and $0.4$ (Figs.~\ref{fig:vorticity_comparison500}b, \ref{fig:vorticity_comparison500}g and \ref{fig:vorticity_comparison500}c, \ref{fig:vorticity_comparison500}h), the wake transitions to a reverse von K\'arm\'an vortex street in all cases. This transition coincides with increasing values of $\overline{C_T}$. With higher $f^*$, the strength of the coherent structures also intensifies. For the actively pitching tail at $f^* = 0.5$ and $f^* = 0.6$, the wake remains reverse von K\'arm\'an, with the vortices becoming progressively stronger and persisting further downstream (Figs.~\ref{fig:vorticity_comparison500}d–\ref{fig:vorticity_comparison500}e). It is consistent with the steady increase in $\overline{C_T}$ observed at higher Strouhal frequencies. 

In contrast, at $f^* = 0.5$ and $f^* = 0.6$, the passively pitching tail produces asymmetric wakes (Figs.~\ref{fig:vorticity_comparison500}i–\ref{fig:vorticity_comparison500}j). These asymmetric reverse von K\'arm\'an vortex streets are generally associated with stronger thrust-producing wakes. Godoy-Diana et al.~\cite{godoy2009model} highlighted that such symmetry breaking occurred at high Strouhal frequencies. This asymmetry implies that the net force generated by the flapping tail is no longer aligned with the mid-plane of the swimmer.

\begin{figure}[ht!]
    \centering
    \includegraphics[width=1\linewidth]{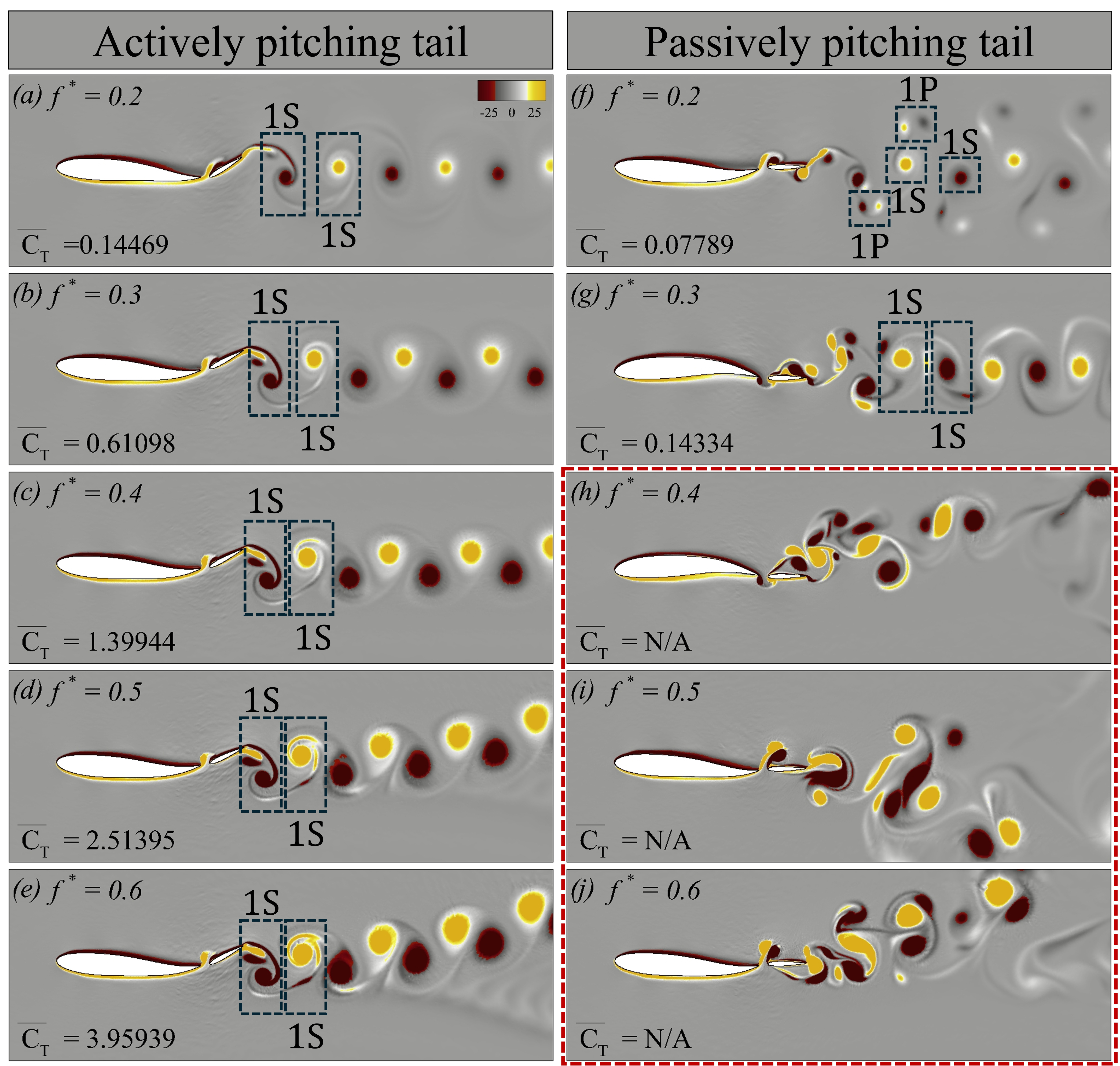}
    \caption{Vorticity contours corresponding to the swimmer with (a-e) an actively pitching tail, and (f-j) a passively pitching tail respectively at Re = $5000$.}
    \label{fig:vorticity_comparison5000}
\end{figure}

Similarly, we now examine the vortex dynamics of both swimmers at Re = $5000$, as shown in Fig.~\ref{fig:vorticity_comparison5000}. The overall layout of this figure is identical to that of Fig.~\ref{fig:vorticity_comparison500}. For the case with a passively pitching tail at $f^* = 0.4$–$0.6$, the results exhibit instabilities, as highlighted by the red periphery in Figs.~\ref{fig:vorticity_comparison5000}h–\ref{fig:vorticity_comparison5000}j.  In Fig.~\ref{fig:vorticity_comparison5000}, $1P$ corresponds to a single vortex pair shed during a half oscillation cycle (one clockwise rotating vortex paired with a counter-clockwise rotating vortex).

Figures~\ref{fig:vorticity_comparison5000}a and \ref{fig:vorticity_comparison5000}f show the wake topology of the actively and passively pitching tail configurations, respectively, at $f^* = 0.2$. For the case with an active tail, the wake resembles that observed at Re = $500$, whereas the passive tail exhibits a $2S$–$2P$ shedding pattern. As the $f^*$ increases to $0.3$, the actively pitching tail maintains the same topology, while the passive tail develops a reverse von K\'arm\'an vortex street characterized by a $2S$ wake pattern. Although the wake corresponds to a reverse von K\'arm\'an street, minor instabilities begin to appear, as shown in Fig.~\ref{fig:vorticity_comparison5000}g. For the actively pitching tail, an increasing Strouhal frequency continues to increase thrust, consistent with the trend observed at Re = $500$. However, from $f^* = 0.4$–$0.6$, the wake becomes increasingly asymmetric, with the asymmetry becoming more pronounced at ma higher $f^*$ (Figs.~\ref{fig:vorticity_comparison5000}c–\ref{fig:vorticity_comparison5000}e). On the other hand, for the passive tail, the red-outlined region (Figs.~\ref{fig:vorticity_comparison5000}h–\ref{fig:vorticity_comparison5000}j) highlights the growing instabilities in the wake as the Strouhal frequency increases. 

%%%%%%%%%%%%%%%%%%%%%%%%%%%%%%%%%%%%%%%%%%%%
%\subsection{Vortex dynamics for a half oscillation cycle}

\begin{figure}[ht!]
    \centering
    \includegraphics[width=1\linewidth]{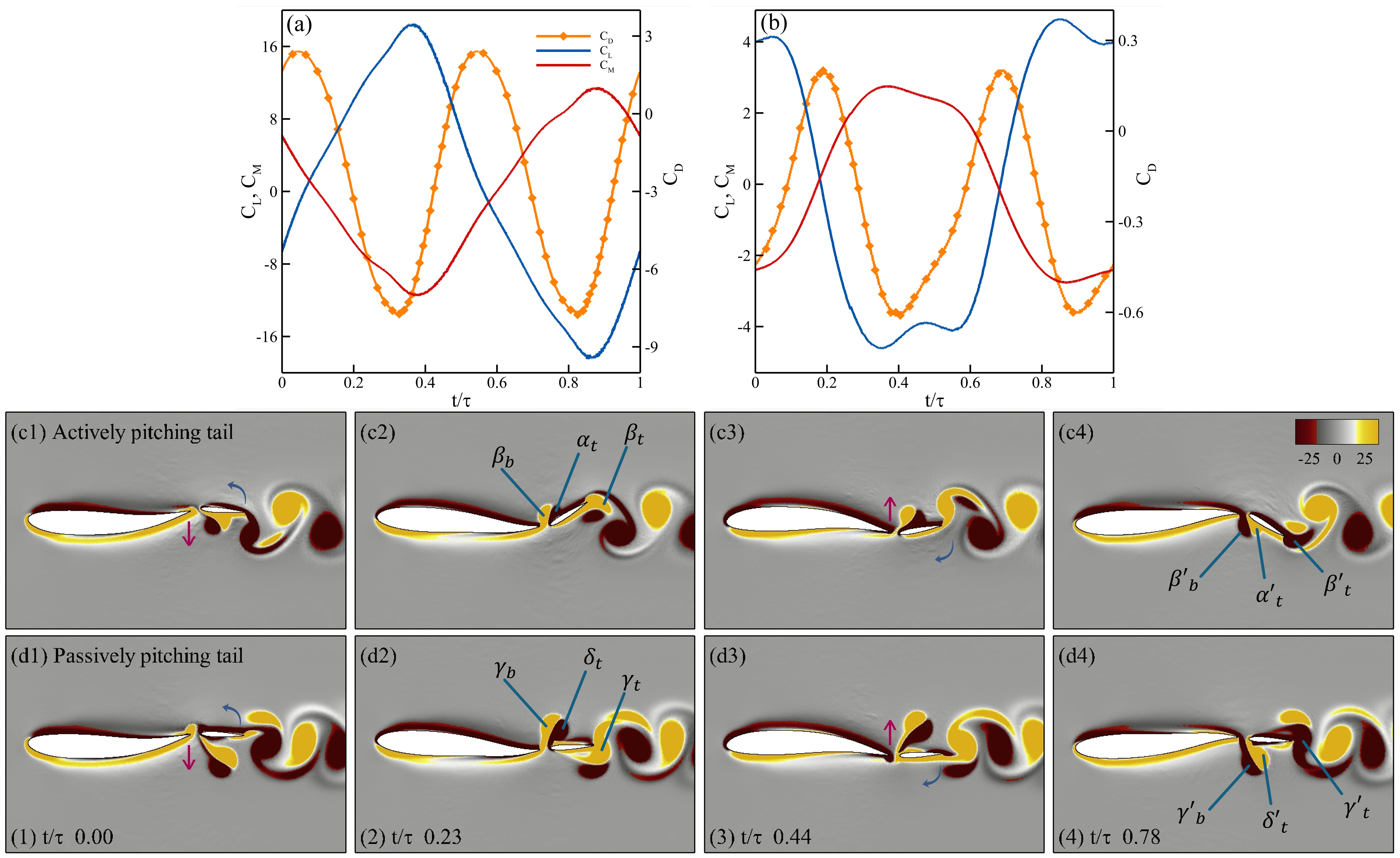}
    \caption{Vortex dynamics of actively and passively pitching tails at Re $500$ and $f^* = 0.6$. Panels (a) and (b) show the variations of force coefficients $C_D$, $C_L$, and $C_M$ over one oscillation cycle for the actively and passively pitching tails, respectively. Panels (c) and (d) depict four characteristic snapshots (1-4) of the flow field for each case, respectively, visualized using contours of $z$-vorticity corresponding to the same cycle.}
    \label{fig:partA}
\end{figure}

From the earlier analysis of the power ratio at $f^* = 0.6$ and Re = $500$, the passively pitching tail ($J^* = 0.125, \; \zeta = 1.0$) produce more than twice the power ratio of than an actively pitching tail at the same $f^*$. To better understand the effect of the flow on the tail, we examine the vortex dynamics of both configurations, shown side by side in Figs~\ref{fig:partA}. Figures~\ref{fig:partA}(a) and \ref{fig:partA}(b) present the drag ($C_D$), lift ($C_L$), and moment ($C_M$) coefficients over one oscillation cycle for the cases with an actively pitching tail and a passively pitching tail, respectively. The corresponding $z$-vorticity fields are displayed in Figs~\ref{fig:partA}(c) and \ref{fig:partA}(d) at four time instants, capturing the motion as the body undulates from its uppermost position, descending to the lowest point, and ascending back to complete the full cycle. In both cases, the undulation of the body and the heaving of the tail remain in phase.

In Fig.~\ref{fig:partA}(c), panel c$1$ shows the tail at the start of its downward heaving stroke. By panel c$2$, a leading-edge vortex ($\alpha_t$) forms on the pressure side of the tail, producing positive lift and negative drag (i.e., positive thrust). Simultaneously, the tail pitches downward, generating a trailing-edge vortex ($\beta_t$), while the body’s undulation produces another vortex at the trailing edge of the body ($\beta_b$). Due to the effect of $\beta_b$, leading to an early formation of $\alpha_t$ which results in the reduction of the pressure on the pressure side. This increase in the pressure differential results in increase in the thrust observed in Fig~\ref{fig:partA}(a), similar to the observation drawn by Akhtar et al \cite{akhtar2005biologically}. As the tail completes its downward stroke (panel c$3$), it begins to pitch upward and $\beta_t$ starts to detach from the trailing edge. During the subsequent upward stroke (panel c$4$), $\alpha_t$ constructively interacts with the newly formed trailing-edge vortex ($\beta'_t$), while a new leading-edge vortex develops on the suction side, resulting in negative lift and negative drag (positive thrust) due to the decrease in the pressure on suction side.

\begin{figure}[ht!]
    \centering
    \includegraphics[width=1\linewidth]{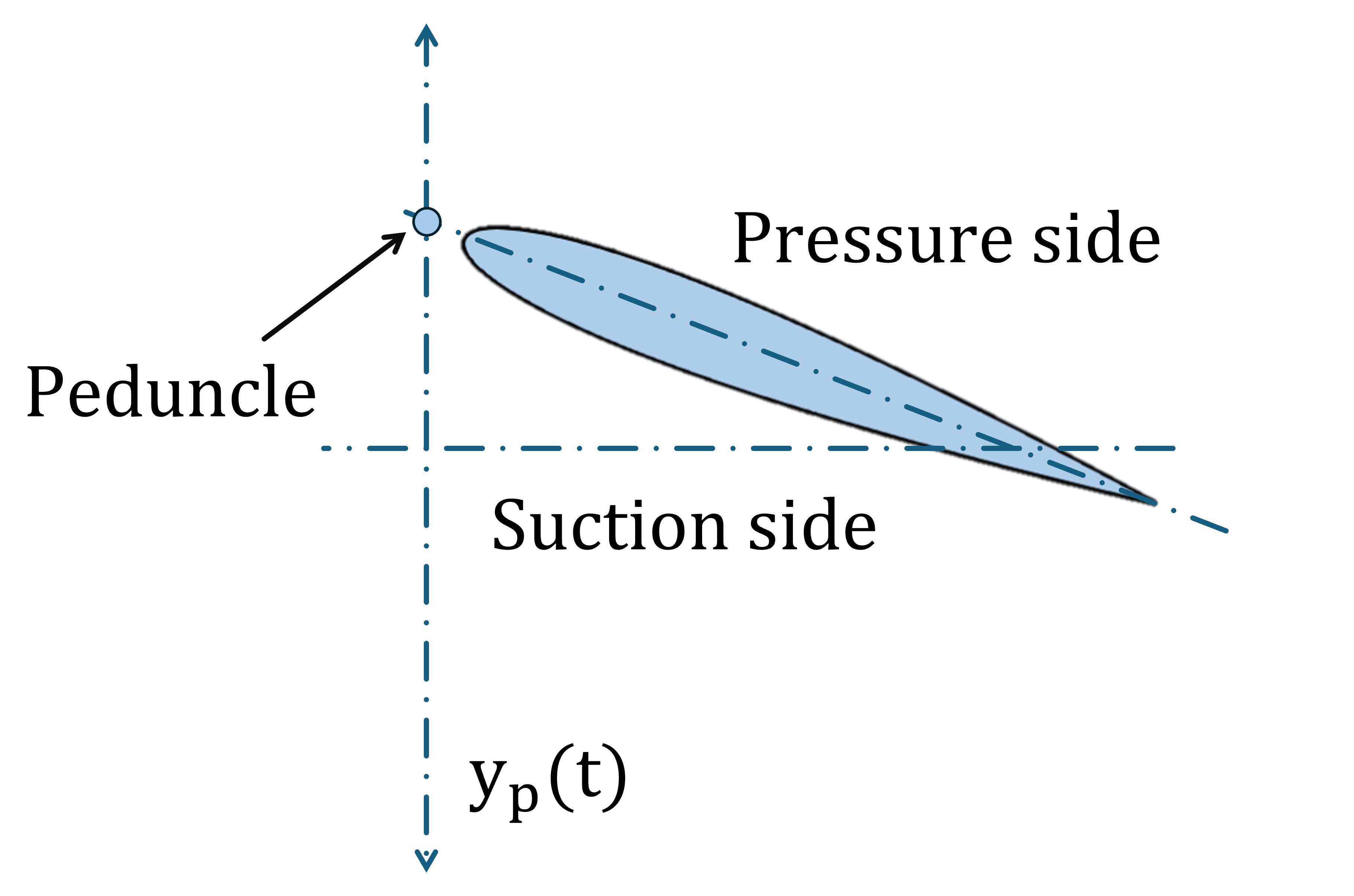}
    \caption{Pressure and Suction sides of the tail}
    \label{fig:tailSchematic}
\end{figure}

Figures.~\ref{fig:partA}(d1) to \ref{fig:partA}(d4), shows the four instances of the passively pitching tail from panel d$1$ to d$4$, respectively. As the tail initiates its downstroke, the lift coefficient decreases (creating downward force) which is in the same direction of the heaving displacement contributing to less power expenditure by the tail. The moment coefficient increases as the tail heaves downward, which leads to the counterclockwise pitching of the tail about its peduncle. As the tail moves downward there is a formation of a larger vortex at the leading edge of the tail ($\delta_t$) and the trailing edge of the body ($\gamma_b$). The previously shed vortex from the leading edge of the tail constructively interferes with the trailing edge vortex ($\gamma_t$) as seen in panel d$2$, increasing the strength of this trailing edge vortex. As seen from panel d$3$, due to the smaller angle of attack of the tail, $\delta_t$ does not stay attached to the boundary layer, which leads to the separation of $\delta_t$ resulting in the formation of a dipole with $\gamma_b$, which is considered detrimental to the production of thrust. When this scenario is compared to the one observed in the actively pitching tail, the formation of the leading edge vortex in both cases leads to the increase in thrust but, with an actively pitching tail, due to the higher angle of attack, the vortex stays attached to the boundary and produces relatively higher thrust. In the same frame, the formation of the  trailing edge vortex ($\gamma'_t$) begins as the tail initiates its upward stroke. In panel d$4$, the dipole which is formed from $\gamma_b$ and $\delta_t$ breaks as $\delta_t$ constructively merges into $\gamma'_t$ which leads to an increase in thrust. Fig.~\ref{fig:tailSchematic} illustrates the pressure side and suction side of the tail at the instant when it is completing the upward heaving stroke and is pitching upward about the peduncle. 

\begin{figure}[ht!]
    \centering
    \includegraphics[width=1\linewidth]{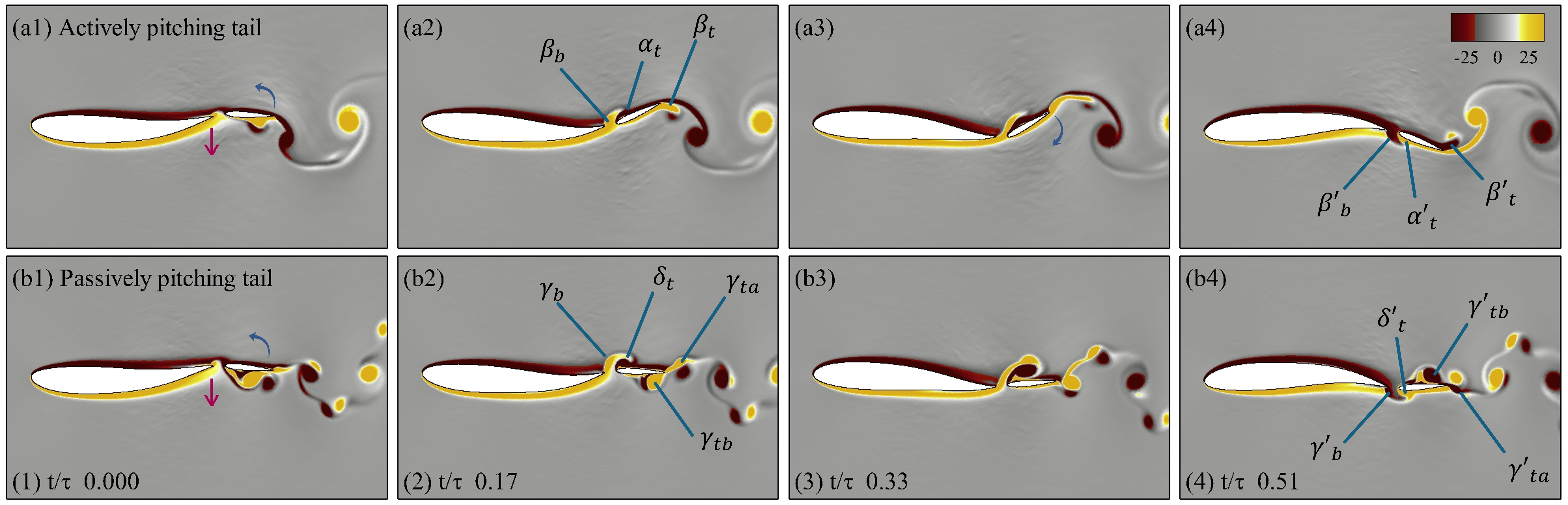}
    \caption{Vortex dynamics of actively and passively pitching tails at Re $5000$ and $f^* = 0.2$. Panels (a) and (b) depict four characteristic snapshots ($1$-$4$) of the flow field for each case, respectively, visualized using contours of $z$-vorticity corresponding to a half oscillation cycle.}
    \label{fig:partB}
\end{figure}

Based on the earlier analysis, at a Reynolds number of Re =$500$, $78\%$ of the passively pitching tail configurations outperformed their actively pitching counterparts in terms of power ratio. In contrast, at a higher Reynolds number of Re $= 5000$, only $38\%$ of the passively pitching tail configurations demonstrated better performance, with most of these cases clustered around a Strouhal number of $0.2$. Notably, the passive tail configuration with parameters $\zeta = 0.50$ and $J^* = 0.375$ achieved nearly $1.5$ times better performance than that of the corresponding active case at the same Strouhal frequency. A comparative analysis of the vortex dynamics over a half oscillation cycle for both cases is presented in Fig.~\ref{fig:partB}, shown in comparison.

Figures~\ref{fig:partB}(a)–\ref{fig:partB}(d) illustrate four key time instances of an actively pitching tail during its downward stroke, corresponding to panels a$_1$–a$_4$, respectively. In panel a$1$, as the tail initiates its downward stroke, a trailing-edge vortex ($\beta_b$) begins to form due to the body's undulation. As the tail continues to move downstroke, a leading-edge vortex ($\alpha_t$) develops on the pressure side of the tail. Simultaneously, as the tail pitches downward about the peduncle, a trailing-edge vortex ($\beta_t$) is formed. A key observation between the actively pitching tail at Re = $500$ (Fig.~\ref{fig:partA}c) and Re =$ 5000$ (Fig.~\ref{fig:partB}a) is that while $\alpha_t$ remains attached to the boundary layer in both cases, at the lower Reynolds number, the vortex structure appears more coherent. Contrarily, at Re $=5000$, the leading-edge vortex is more tightly bound to the boundary layer, contributing to improved thrust generation. In panel a$3$, as the tail approaches the end of its downward stroke, the interaction of $\beta_b$ with $\alpha_t$ produces a jet-like effect, similar to the mechanism described by Gao et al.~\cite{gao2018independent}.  At this instance, the trailing-edge vortex continues to strengthen as the tail's angle of attack reaches its peak. Panel a$4$ captures the moment just after the downward stroke ends. Here, the upward pitching of the tail causes the detachment of the trailing-edge vortex $\beta_t$, while a new vortex $\beta'_t$ is formed on the pressure side. The constructive interaction between $\alpha_t$ and $\beta'_t$ leads to the production of thrust. At this same instant, a new leading edge vortex ($\alpha'_t$) from the tail and a trailing-edge vortex ($\beta'_b$) from the body appears on the suction side. 

Similarly, Fig.~\ref{fig:partB}(b) illustrates the passively pitching tail over a half oscillation cycle as it completes its downward stroke. The four snapshots of this motion are shown in panels b$1$–b$4$. In panel b$1$, as the tail begins its downstroke, a trailing-edge vortex ($\gamma_b$) is generated from the undulating body from the suction side. A core difference between the cases with an actively pitching tail and the passively pitching tail lies in the angle of attack. For the passive tail, the angle of attack does not exceed $6^\circ$, which is significantly smaller than $30^\circ$ observed in the active tail. This reduced angle of attack is consistent across all configurations with the passively pitching tail. However, due to the smaller angle of attack, the leading-edge vortex ($\delta_t$) detaches prematurely from the boundary layer of the tail. As it remains close to the surface, $\delta_t$ subsequently reattaches to the boundary layer, as seen in panel b$3$. Upon reaching the trailing edge of the tail, it sheds as a coherent, clockwise-rotating vortex from the pressure side, relabeled as $\gamma'_{tb}$. Such reattachment and delayed shedding help maintain a favorable pressure distribution along the surface of the tail and promotes the formation of a reverse jet downstream, thereby contributing to the generation of thrust as seen in panel b$4$. In the same panel, we also observe the formation of a new leading-edge vortex ($\delta'_t$) on the tail and a trailing-edge vortex ($\gamma'_b$) from the undulating body on the suction side.

The primary contributor to the higher thrust observed in all actively pitching tail configurations is the large angle of attack and a consistent lag between the heaving and the pitching displacement of the tail, which ensures that the leading-edge vortex formed on the tail remains attached to the boundary layer, thereby sustaining thrust generation. A secondary contributor is the constructive interaction between the leading-edge vortex from the first half of the oscillation cycle and the trailing-edge vortex formed in the second half, which further improves the production of thrust. However, this higher thrust comes at the cost of a reduced power ratio. Contrarily, for the passively pitching tail, the smaller angle of attack causes the leading-edge vortex to detach prematurely from the boundary layer of the tail, which is detrimental for the production of the thrust. Nevertheless, when performance is evaluated in terms of power ratio, the passively pitching tail outperforms its active counterpart at Re = $500$. At Re = $5000$, however, the trend reverses: the actively pitching tail exhibits better performance, particularly at higher Strouhal frequencies, due to the instabilities and inconsistencies observed in the case with a passively pitching tail as discussed earlier.

\section{Conclusions}

An actively pitching tail generally produces higher thrust ($-\overline{C_D}$) than its passive counterpart, whereas a passively pitching tail demonstrates higher power ratio ($\eta$) at Re = $500$, opposite to what is observed at Re = $5000$. Larger pitching amplitudes are beneficial only when achieved through increased Strouhal frequency ($f^*$); when high amplitudes result from smaller inertia ($J^*$) or looser joints (lower $\zeta$), the effect becomes detrimental. At Re = $5000$, actively pitching tail configurations dominate in performance based on power ratio, with only a few passive cases near $f^* = 0.2$ outperforming the active configuration. A consistent trade-off emerges in all cases: configurations producing higher thrust tend to exhibit lower power ratio, while those with lower thrust achieve higher power ratio. One unique case at Re = $5000$ ($\zeta = 0.25$, $J^* = 0.125$) displayed alternating thrust- and drag-dominant behavior with a maximum pitching angle of $25^\circ$, but due to its rarity, it is not examined further. From a physical standpoint, smaller swimmers may benefit from a passively pitching tail, whereas larger swimmers are more likely to find such behavior detrimental and may perform better by allowing the tail to pitch actively.

\section*{Acknowledgment}
MSU Khalid acknowledges the funding support the Natural Sciences and Engineering Research Council of Canada (NSERC) through the Discovery grant program. A. Tarokh also confirms the support from NSERC for this research. The simulations reported in this work were performed on the supercomputing clusters administered and managed by the Digital Research Alliance of Canada. 

\bibliography{references}

%%%%%%%%%%%%%%%%%%%%%%%%%%%%%%%%%%%%%%%%%%%%%%%%%%%%%%%%%%%%%%%%%%%%%%%%%%%%%%%
\end{document}